\documentclass{astlb}
\usepackage{graphicx}
\usepackage{floatrow}
\usepackage{longtable}

\usepackage[hyperfootnotes=false,draft]{hyperref}
\usepackage{changes}

\usepackage{amsmath,amssymb}

\def\arcsec{\hbox{$^{\prime\prime}$}}
\def\arcmin{\hbox{$^{\prime}$}}

\def\ps1{{Pan-STARRS1}}

\def\doubleline{\vskip 3pt\hrule \vskip 1.5pt \hrule \vskip 5pt}

\def\smfigure#1#2#3{
  \begin{minipage}{1.0\columnwidth}
    \begin{minipage}{0.049\columnwidth}
      \rotatebox{90}{\footnotesize\phantom{0000}#3}
    \end{minipage}
    \begin{minipage}{0.9\columnwidth}
     \includegraphics[viewport=40 188 556 678,width=0.97\columnwidth]{#1}
      \centerline{\footnotesize #2}
    \end{minipage}

    \vskip 3pt
    ~
  \end{minipage}
}

\def\smfiguresmall#1#2#3{
  \begin{minipage}{0.485\textwidth}
    \begin{minipage}{0.049\columnwidth}
      \rotatebox{90}{\footnotesize\phantom{0000}#3}
    \end{minipage}
    \begin{minipage}{0.90\columnwidth}
     \includegraphics[viewport=30 188 556 500,width=0.97\columnwidth]{#1}
      \centerline{\footnotesize #2}
    \end{minipage}
  \end{minipage}
}

\begin{document}

\journalinfo{2023}{49}{11}{599}[620]

\title{Optical Identification and Spectroscopic Redshift Measurements of\\
216 Galaxy Clusters from the SRG/eROSITA All-Sky Survey}

\author{I. A. Zaznobin\address{1,2}\email{zaznobin@cosmos.ru},
  R.~A.~Burenin\address{1,2},
  A.~A.~Belinski\address{2},
  I.~F.~Bikmaev\address{2,3},
  M.~R.~Gilfanov\address{1,5},
  A.~V.~Dodin\address{2},
  S.~N.~Dodonov\address{6},
  M.~V.~Eselevich\address{7},
  S.~F.~Zheloukhov\address{2},
  E.~N.~Irtuganov\address{3},
  S.~S.~Kotov\address{6},
  R.~A.~Krivonos\address{1},
  N.~S.~Lyskova\address{1},
  E.~A.~Malygin\address{6},
  N.~A.~Maslennikova\address{2},
  P.~S.~Medvedev\address{1},
  A.~V.~Meshcheryakov\address{1},
  A.~V.~Moiseev\address{1,6},
  D.~V.~Oparin\address{6}, 
  S.~A.~Potanin\address{2,8},
  K.~A.~Postnov\address{2},
  S.~Yu.~Sazonov\address{1},
  B.~S.~Safonov\address{2},  
  N.~A.~Sakhibullin\address{3,4},
  A.~A.~Starobinsky\address{9},
  M.~V.~Suslikov\address{2,3},
  R.~A.~Sunyaev\address{1,5},
  A.~M.~Tatarnikov\address{2,8},
  G.~S.~Uskov\address{1},
  R.~I.~Uklein\address{6},
  I.~I.~Khabibullin\address{1,5},
  I.~M.~Khamitov\address{3,4},
  G.~A.~Khorunzhev\address{1},
  E.~M.~Churazov\address{1,5},
  E.~S.~Shablovinskaya\address{6},
  N.~I.~Shatsky\address{2}.  
  \addresstext{1}{Space Research Institute, Russian Academy of Sciences, Moscow, Russia}
  \addresstext{2}{Sternberg Astronomical Institute, Moscow State University, Moscow, Russia}
  \addresstext{3}{Kazan Federal University, Kazan, Russia}
  \addresstext{4}{Academy of Sciences of Tatarstan, Kazan, Russia}
  \addresstext{5}{Max Planck Institut f\"{u}r Astrophysik, Garching, Germany}
  \addresstext{6}{Special Astrophysical Observatory, Russian Academy of Sciences, Nizhnii Arkhyz, Russia}
  \addresstext{7}{Institute of Solar–Terrestrial Physics, Russian Academy of Sciences, Siberian Branch, Irkutsk, Russia}
  \addresstext{8}{Faculty of Physics, Moscow State University, Moscow, Russia}
  \addresstext{9}{Landau Institute for Theoretical Physics, Russian Academy of Sciences, Chernogolovka, Russia}
  }
  
\submitted{}

\begin{abstract}
We present the results of the optical identification and spectroscopic redshift measurements of 216 galaxy clusters detected in the SRG/eROSITA all-sky X-ray survey. The spectroscopic observations were performed in 2020--2023 with the 6-m BTA telescope at the Special Astrophysical Observatory of the Russian Academy of Sciences, the 2.5-m telescope at the Caucasus Mountain Observatory of the Sternberg Astronomical Institute of the Moscow State University, the 1.6-m AZT-33IK telescope at the Sayan Solar Observatory of the Institute of Solar–Terrestrial Physics of the Siberian Branch of the Russian Academy of Sciences, and the 1.5-m Russian–Turkish telescope (RTT-150) at the T\"{U}B\.{I}TAK Observatory. For all of the galaxy clusters presented here the spectroscopic redshift measurements have been obtained for the first time. Of these, 139 galaxy clusters have been detected for the first time in the SRG/eROSITA survey and 22 galaxy clusters are at redshifts $z_{\rm spec} \gtrsim 0.7$, including three at $z_{\rm spec} \gtrsim 1$. Deep direct images with the \textit{rizJK} filters have also been obtained for four distant galaxy clusters at $z_{\rm spec} > 0.7$. For these observations the most massive clusters are selected. Therefore, most of the galaxy clusters presented here most likely will be included in the cosmological samples of galaxy clusters from the SRG/eROSITA survey.

\end{abstract}

\section{INTRODUCTION}

Studying large samples of massive galaxy clusters detected, among others, through microwave and X-ray surveys to establish constraints on the basic parameters of cosmological models for the Universe \citep[see, e.g.,][]{av09cosm,PSZcosm13,PSZ2cosm}. The Spectrum–Roentgen–Gamma (SRG) space observatory \citep{srg} with the onboard grazing-incidence ART-XC \citep{art} and eROSITA \citep{ero} telescopes was launched in July 2019. In December 2019 the first of the eight planned all-sky surveys has started. Four full all-sky surveys have been completed to date. It is expected that upon completion of the eight full all-sky surveys, the achieved survey depth will be enough to detect all galaxy clusters with masses higher than $M_{500} \sim 3 \times 10^{14} M_{\odot}$ in the observable Universe \citep[see, e.g.,][]{churazov15}.

Our group has previously performed the optical identifications and spectroscopic redshift measurements of massive galaxy clusters identified with Sunyaev–Zeldovich sources \citep{PSZ_RTT150,PSZ1Addendum,PSZ_Canary,vorobiev16,br18hz,zazn19,zazn20,zazn21}. In \cite{zazn21} it is shown that most of these galaxy clusters were already detected in the SRG/eROSITA all-sky survey. The observations of galaxy clusters from the Lockman Hole survey have already been carried out previously \citep{zazn21lh}.

After the completion of the first SRG/eROSITA all-sky survey in June 2020, we initiated a program of optical identifications and spectroscopic redshift measurements of the most massive galaxy clusters in the survey. Spectroscopic redshift measurements for galaxy clusters are the main goal of this program. The program of observations was gradually supplemented with increasing number of completed full all-sky surveys. We also initiated a program of deep photometric observations of candidates for distant galaxy clusters and their photometric and spectroscopic redshift measurements.

At present, the DESI spectroscopic sky survey is being conducted \citep{desiedr}. The first year of the five-year DESI survey was completed on June 13, 2022. It is expected that the spectra of about 30 million galaxies and quasar will be taken as a result of three years of the survey. The redshifts for many of the brightest galaxies of the clusters detected in the SRG/eROSITA survey will probably be measured in this survey. However, the observations within our program are still needed to produce the cosmological samples of galaxy clusters from the SRG/eROSITA survey as quickly as possible and to obtain the cosmological constraints from these data.

We present the results of our optical identification and spectroscopic redshift measurements of the most massive galaxy clusters. We show the results of obtaining deep direct images for some distant clusters. In three years from June 2020 to September 2023 we measured the redshifts of 216 galaxy clusters and obtained deep direct images for several most distant galaxy clusters at $z > 0.7$ detected in the SRG/eROSITA survey. The results of our redshift measurements and our estimates of the masses $M_{500}$ for 12 most massive galaxy clusters have already been published previously by our research group \citep{br21,br22}.

\section{THE SAMPLE OF OBJECTS}

The optical identification of galaxy clusters was carried out among the extended X-ray sources detected as a result of the SRG/eROSITA all-sky survey in the half of the sky ($0^\circ < l < 180^\circ$) for the data processing in which the Russian scientists are responsible. The search for extended X-ray sources was carried out using a wavelet decomposition of X-ray images \citep{av98} and the ermldet software from the eSASS package. Some sources can be identified with known galaxy clusters for which the spectroscopic redshifts have been measured previously. The NASA extragalactic database (NED\footnote{https://ned.ipac.caltech.edu/}) was used to identify the known galaxy clusters. For a significant fraction of massive galaxy clusters the spectroscopic redshift measurements have been obtained previously, mostly within the Sloan Digital Sky Survey \citep[SDSS,][]{sdssdr16}.

The publicly accessible data from large optical and near-infrared sky surveys were used for the optical identification. We used data from Pan-STARRS1 \citep{ps1}, DESI LIS \citep{desi}, and the WISE all-sky survey \citep{wise} in the 3.4~$\mu$m band. We used the WISE forced photometry performed using the coordinates of galaxies from Pan-STARRS1 \citep{wise_forsed}. The procedure for the optical identification of galaxy clusters was discussed previously in the papers of our group \citep{br18hz,zazn19,zazn20,zazn21,zazn21lh,br21,br22}. The photometric redshift estimate from the red-sequence color of cluster galaxies was given for each galaxy cluster \citep{br17,br22}.

As a result, we obtained a sample of several thousand galaxy clusters. Obviously, the observing time available to our group is not enough to carry out spectroscopic redshift measurements in a reasonable time for several years for such a large number of galaxy clusters. Therefore, for each galaxy cluster we determined the priority index that was taken into account during our observations:
\begin{equation*}
P = 8.7\sqrt{z}+3.3lg(f)+33,
\end{equation*}
where $z$ is the photometric redshift estimate for the cluster and $f$ is the X-ray ﬂux. The highest priority goals for our observations have the highest priority index. This index is defined in such a way that the most massive clusters had the highest priority and, in addition, that the clusters of one mass at different redshifts had approximately the same priority. We carried out the observations of predominantly galaxy clusters whose priority index was higher than $-3$\ldots$-4$. This strategy of observations allows complete samples of the most massive clusters with masses higher than $3\times 10^{14} M_\odot$ to be obtained.

\section{OBSERVATIONS}

\subsection{Organization of Observations}

The spectroscopic observations of galaxy clusters have been carried out since 2020 with the 6-m BTA telescope at the Special Astrophysical Observatory of the Russian Academy of Sciences (SAO RAS), the 1.6-m AZT-33IK telescope at the Sayan Solar Observatory of the Institute of Solar–Terrestrial Physics of the Siberian Branch of the Russian Academy of Sciences, and the 1.5-m Russian–Turkish telescope (RTT-150) at the T\"{U}B\.{I}TAK Observatory. Since January 2020 the observations have also been carried out with the 2.5-m telescope at the Caucasus Mountain Observatory (CMO) of the Sternberg Astronomical Institute of the Moscow State University (RC2500). The observations with the BTA and 2.5-m CMO telescopes are carried out within the optical observations program of SRG ground support for the optical identification of the most massive distant galaxy clusters from the SRG/eROSITA all-sky survey.

The objects were selected by the observers for the subsequent observations from the lists accessible to the observers in the TRITON system (the table of X-ray sources for which optical observations are required)\footnote{https://www.srg.cosmos.ru/triton/welcome}. The system was created as a site at which data for the optical observations of X-ray sources detected in the SRG survey are provided. The main task of the system is the collection, storage, and systematization of data on the optical observations of SRG objects and the results of their processing. The system allows the observers to trace the observational work at various telescopes being carried out or planned in real time. This allows the repeated or simultaneous observations of objects to be avoided, which is very important in the case of simultaneous observations at several telescopes.

At the BTA and 2.5-m CMO telescopes we carried out the observations of, as a rule, the most distant and faintest objects, whose observations are difficult to perform at the AZT-33IK and RTT-150 telescopes. At the AZT-33IK and RTT-150 telescopes we carried out the observations of nearby galaxy clusters with a photometric redshift estimate $z_{\rm phot} \lesssim 0.55$. The diffraction gratings for each cluster were selected in such a way that, given their photometric redshift estimates, some main spectral features of elliptical galaxies, such as the 4000~\AA\ break and the G line, entered the spectral range of the gratings.

\subsection{Observations and Instruments}

Low- and medium-resolution long-slit spectrographs were used to obtain the spectroscopic images. The position angle of the spectrograph slit for each object was selected individually in such a way that the light from as many brightest red-sequence galaxies as possible fell into the spectrograph slit. For each galaxy cluster we determined one slit configuration: the center coordinates and the position angle of the spectrograph slit. The slit width was selected separately for each spectrograph in the range from 1\arcsec\ to 2\arcsec\, depending on the seeing. If more than two red sequence galaxies fell on the spectrograph slit, then we set a slit about 2\arcsec\ in width irrespective of the instrument being used.

The observations at the BTA telescope were carried out with the SCORPIO \citep{scorpio05} and SCORPIO-2 \citep{scorpio11} spectrographs. We used the VPHG940@600, VPHG1026@735, and VPHG1200@860 grisms during the observations with the SCORPIO-2 spectrographs and the VPHG550G grism during the observations with the SCORPIO spectrograph. To obtain the spectra of galaxies at SCORPIO-2 with photometric redshift estimates $z_{\rm phot}\lesssim 0.75$, $0.75 \lesssim z_{\rm phot} \lesssim 1.00$, and $z_{\rm phot} \gtrsim 1$, we used the VPHG940@600, VPHG1026@735, and VPHG1200@860 grisms, respectively. The instruments and the grisms being used by us are described in detail at the site of the Laboratory of Spectroscopy and Photometry for Extragalactic Objects at SAO RAS\footnote{https://www.sao.ru/hq/lsfvo/devices\_rus.html}. No spectroscopy of distant galaxy clusters was performed with the SCORPIO spectrograph, since a CCD array with a strong fringe effect in the near infrared is installed at this spectrograph as a detector.

The observations with the 2.5-m telescope at the Caucasus Mountain Observatory of the Sternberg Astronomical Institute of the Moscow State University were carried out using the Transient Double beam Spectrograph \citep[TDS,][]{Potanin20}\footnote{http://lnfm1.sai.msu.ru/kgo/instruments/tds}. The spectral range of the instrument is 3600\,--\,7500\,\AA, and the resolution is 1300\,--\,2500 at a slit width of 1\arcsec. The observing and spectral image processing procedures are described in more detail in \cite{Dodin21}.

The observations at the AZT-33IK telescope were carried out with the low- and medium-resolution ADAM spectrograph \citep{adam16,azt33ik16}. To obtain the spectroscopic images, we used the grism with the volume phase holographic grating VPHG600G with 600 lines per millimeter, which allows one to obtain a spectrum in the range 3600\,--\,7250\,\AA\ with a spectral resolution (FWHM) of 4.3\,\AA\ for a slit width of 2\arcsec. The position angle of the spectrograph slit is $0^{\circ}$. All our spectroscopic images were obtained using a long slit of width 2\arcsec.

The RTT-150 observations were carried out with the TFOSC\footnote{http://hea.iki.rssi.ru/rtt150/en/index.php?page=tfosc} spectrograph. We used a diffraction grating with the spectral range 3800\,--\,8900\,\AA\ and a resolution $R\approx 500$. The position angle of the spectrograph slit is $90^{\circ}$. All our spectroscopic images were obtained using a long slit of width 1.8 and 2.4\arcsec.

On each night at dusk and dawn twilight time we carried out the observations of spectrophotometric standards. The list of standards being used is given at the site of the European Southern Observatory\footnote{https://www.eso.org/sci/observing/tools/standards.html}. The data from BTA, AZT-33IK, and RTT-150 were processed using our own software, the IRAF\footnote{https://iraf-community.github.io/} software package, and its integration in the Python3 programming language (the PyRAF\footnote{http://stsdas.stsci.edu/pyraf/doc.old/pyraf\_tutorial/} library).

For the most distant galaxy clusters at redshifts $z \gtrsim 0.7$ we obtained deep direct images at AZT-33IK, RTT-150, and BTA using the \textit{riz} filters of the SDSS system and the \textit{J} filter during the observations at the 2.5-m CMO telescope. We carried out the observations at AZT-33IK with the Andor iKon-M 934 BR DD camera and at the 2.5-m CMO telescope with the AstroNIRCam infrared spectrograph camera \citep{astronircam, Tatarnikov}. At all telescopes the observations were carried out at seeing not poorer than 1.5\arcsec. The exposure times at BTA, AZT-33IK, and RTT-150 did not exceed 90, 120, and 600~s, respectively. After each exposure, the image center was shifted in a random direction in right ascension or declination by $\approx 10$\arcsec\,--\,$15$\arcsec.

\subsection{Correction for the Atmospheric Oxygen Absorption}

For some of the distant galaxies observed at BTA and located at redshift $z\sim0.9$ the part of the spectrum with the Balmer jump (4000\,\AA) falls into the region of the strong atmospheric oxygen absorption band (7580--7700\,\AA). This makes it difficult to use this spectral feature in the methods of redshift estimates for such sources. This is particularly important in the estimates of the distances to galaxy clusters, where the spectra of faint late-type galaxies without emission lines are investigated. The observations of spectrophotometric standards of early spectral types allow one to estimate the atmospheric oxygen absorption in the region of the band 7580--7700\,\AA\ and to reconstruct the spectra of faint sources.

For this purpose, the standards were observed in sky regions close to the objects at the same zenith distance. However, for various reasons, the standards and objects can be observed at different airmasses. In this case, provided that the observations were carried out in spectrophotometric conditions, the atmospheric absorption in this oxygen band is estimated as 
\begin{equation*}
    P(\lambda)= \Bigr(\frac {I(\lambda)} {I_0(\lambda)} \Bigl)^{\sqrt{\frac {M_{\rm obj}} {M_{\rm std}} }},
\end{equation*}
where $ I(\lambda) $ is the intensity being recorded, $ I_0(\lambda) $ is the intensity without the atmospheric oxygen absorption, $ M_{\rm obj} $ is the airmass of the object being observed, and $ M_{\rm std} $ is the airmass of the standard being observed. The square root was taken into account in view of the so-called square root law for significantly saturated telluric lines, which the oxygen absorption band is.

We mostly carried out the observations of standards that are stars of early spectral types and have a smooth profile without features in the segment of the spectrum under consideration. In the same place the change in the distribution of Rayleigh scattering together with the sensitivity of the receiving equipment can also be assumed, to a first approximation, to be linear. Then, $I_0(\lambda)$ is the linear fit to the continuum in the segments free from telluric absorption lines.

\begin{figure}
  \centering

  \includegraphics[width=\columnwidth]{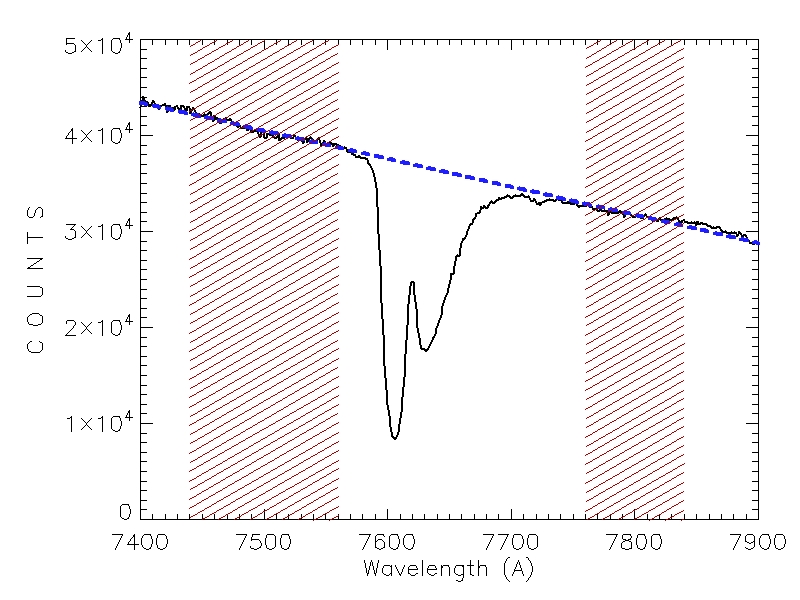}

  \hspace{1pt}
  \caption{Instrumental spectrum of the spectrophotometric standard BD+25d4655 in the region of the $O_2$ absorption band (7580--7700\,\AA).}
  \label{fig:atm}
  
\end{figure}

We selected two segments, 7440$<$$\lambda$$<$7560\,\AA\ and 7760$<$\,$\lambda$\,$<$\,7840\,\AA, to fit the continuum in the region of the spectrum under consideration. figure~\ref{fig:atm} shows examples of the instrumental spectrum for the spectrophotometric standard BD+25d4655, the linear fit to the continuum (blue dashed line), and the regions free from the inﬂuence of telluric lines based on which the spectrum was fitted.

\section{RESULTS OF OBSERVATIONS}

\subsection{Data}

\begin{figure*}
  \centering

  SRGe\,J105036.9$+$455126, $z=0.8747$
  \medskip
  
  \includegraphics[width=0.31\columnwidth]{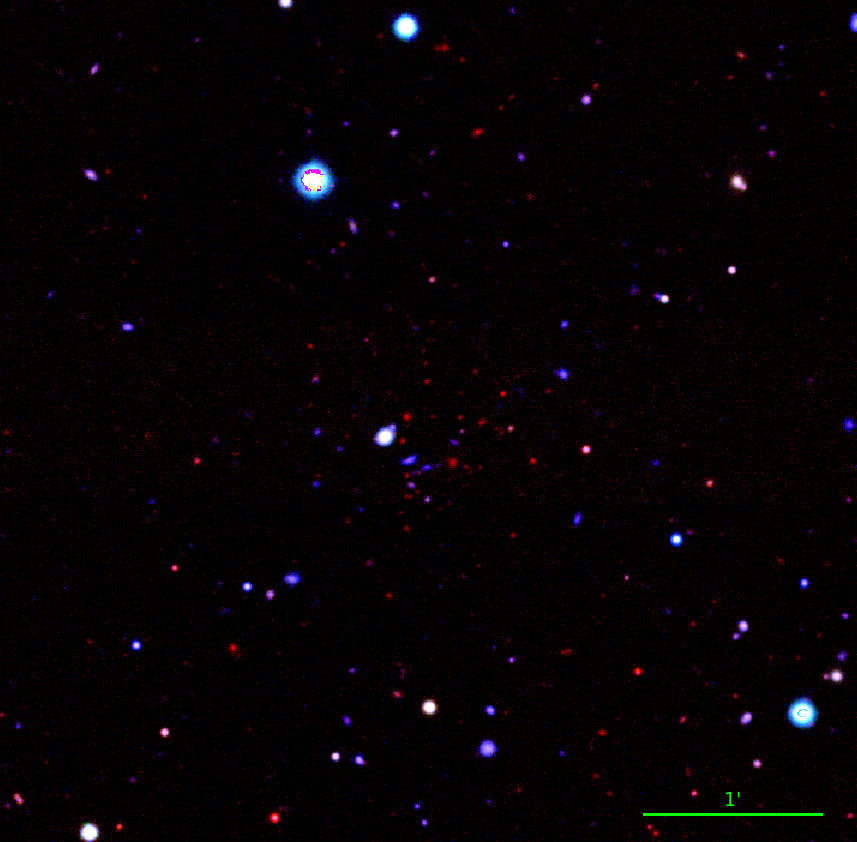} 
  \includegraphics[width=0.31\columnwidth]{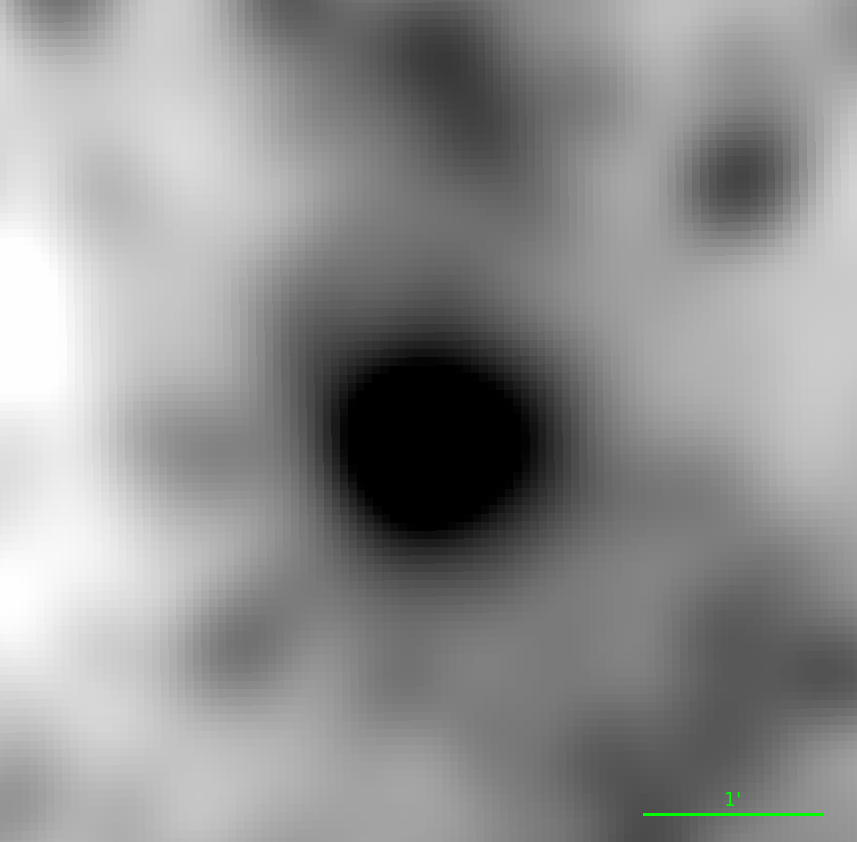}
  \includegraphics[width=0.31\columnwidth]{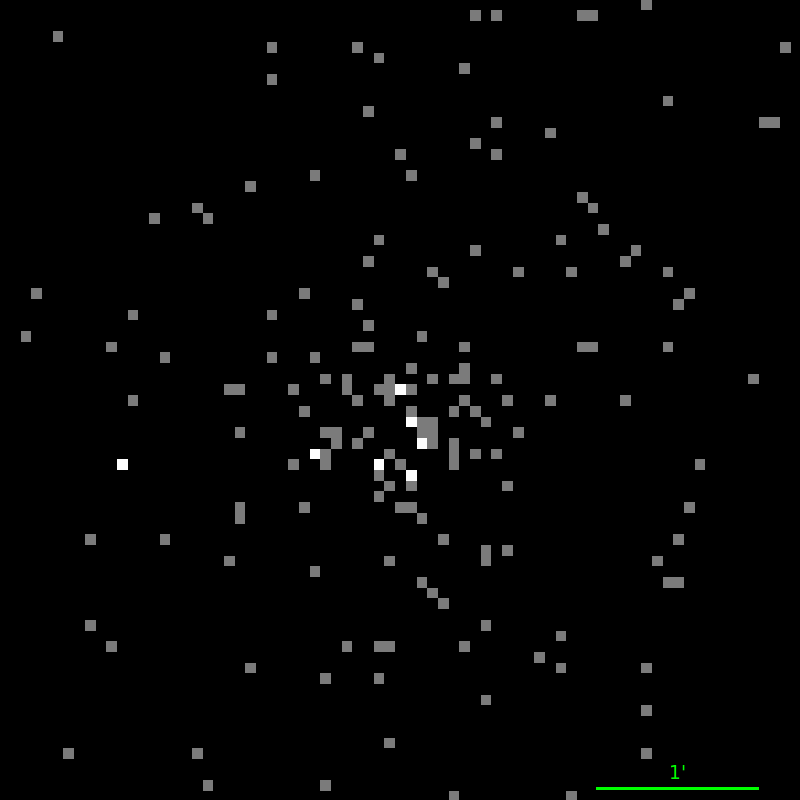}
  \bigskip

  SRGe\,J144035.3$+$661630, $z=0.699$
  \medskip
  
  \includegraphics[width=0.31\columnwidth]{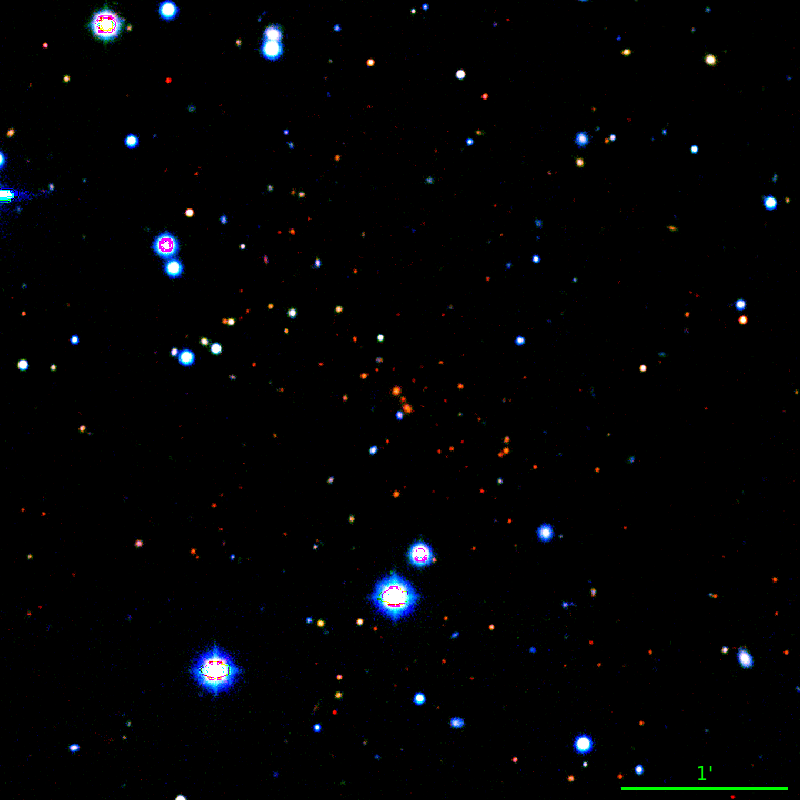} 
  \includegraphics[width=0.31\columnwidth]{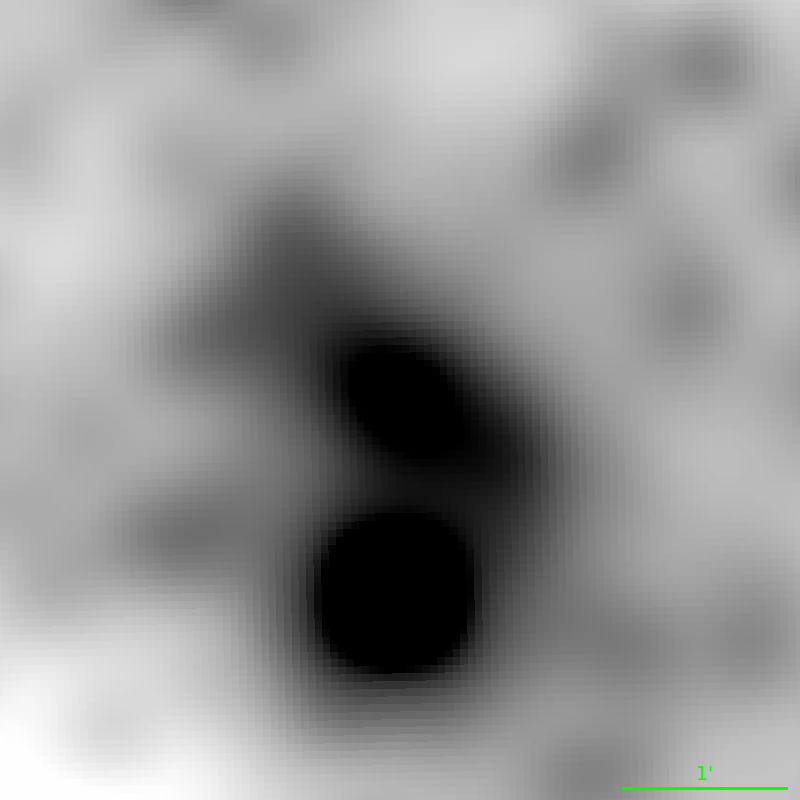}
  \includegraphics[width=0.31\columnwidth]{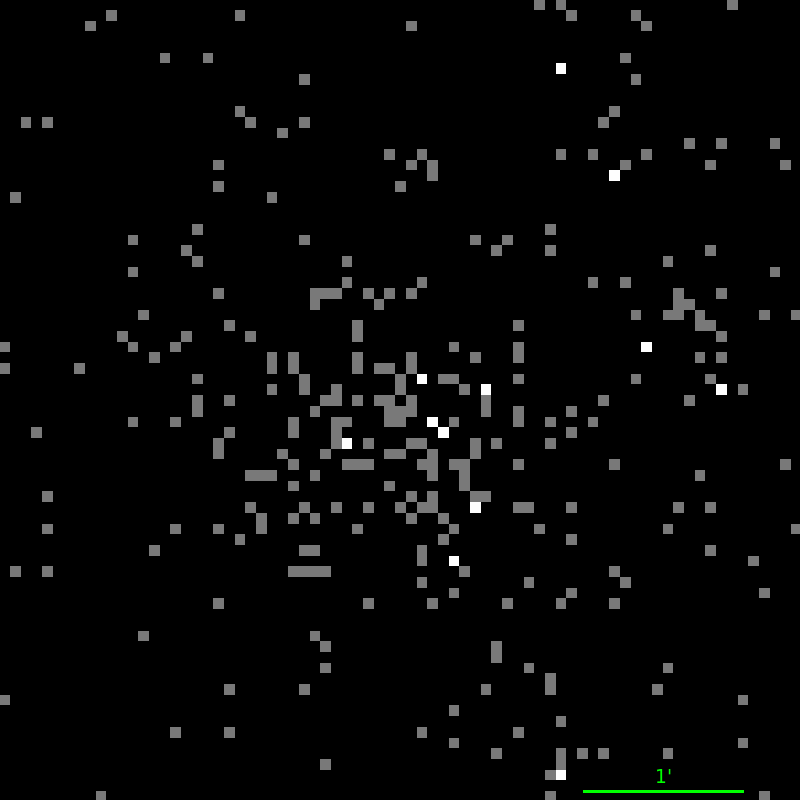}
  \bigskip

  SRGe\,J215157.4$+$111248, $z=0.919$
  \medskip
  
  \includegraphics[width=0.31\columnwidth]{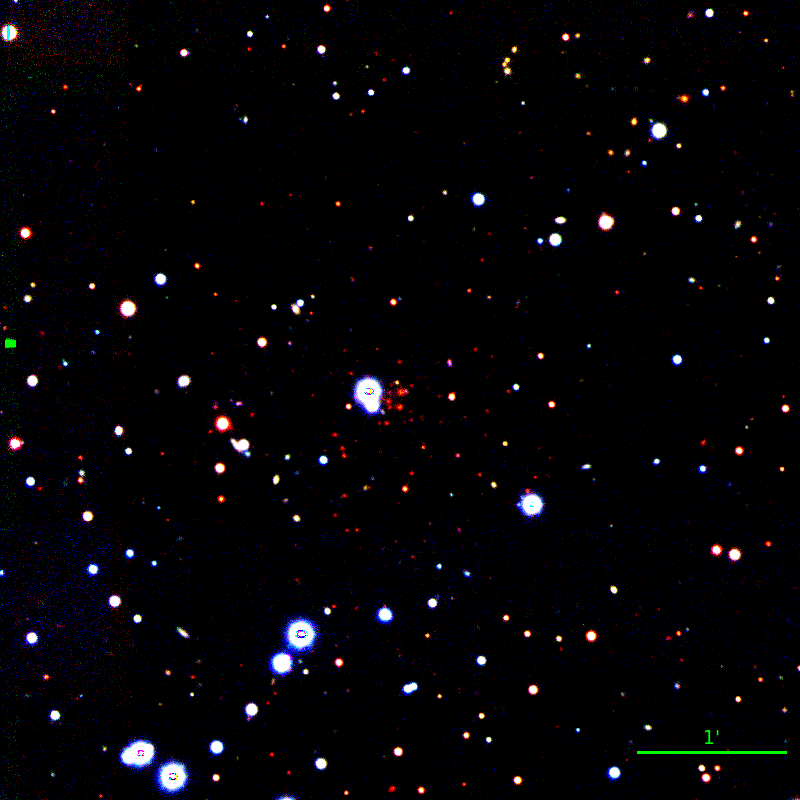} 
  \includegraphics[width=0.31\columnwidth]{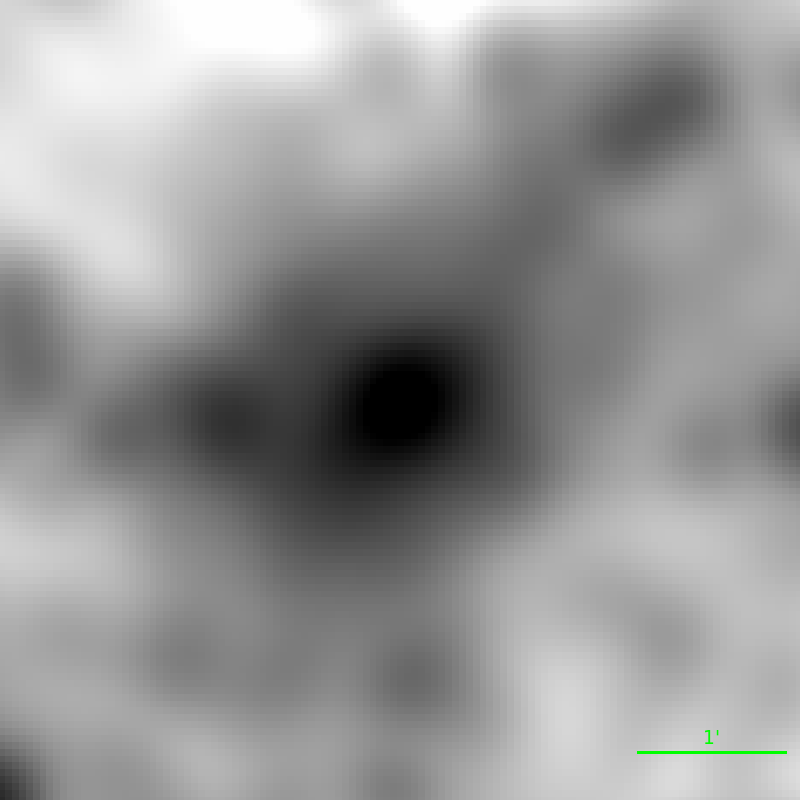}
  \includegraphics[width=0.31\columnwidth]{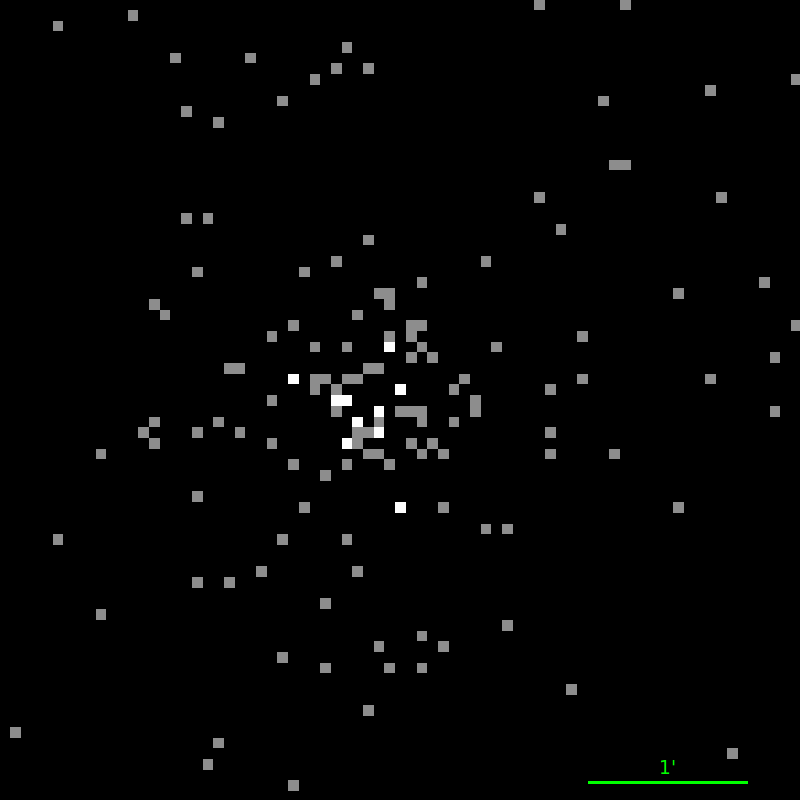}
  \bigskip
  
  \bigskip
  
  \caption{Left: pseudo-color DESI LIS images of the galaxy cluster ﬁelds in the \emph{zrg} (RGB) ﬁlters. Center: WISE images in the 3.4~$\mu$m band cleaned of stars and convolved with a $\beta$-model of radius 24\arcsec. Right: SRG/eROSITA X-ray images. The image center coincides with the optical cluster center, the sizes of the ﬁelds are $5\arcmin \times 5\arcmin$.
  }
  \label{fig:dir_image}

\end{figure*}

As a result of the observations of galaxy clusters from the SRG/eROSITA survey within our program, by September 2023 we obtain the spectra and measured the redshifts for 216 clusters. All these galaxy clusters previously had no spectroscopically measured redshifts. Of these, 106, 45, 51, and 26 galaxy clusters were observed at BTA, RC2500, AZT-33IK, and RTT-150, respectively. Some of the galaxy clusters were observed at several telescopes, since low-quality spectra were taken due to the weather conditions from which the redshifts of galaxies cannot be measured reliably. In these cases, we repeated the observations at a different telescope. Examples of the optical, infrared, and X-ray images of galaxy clusters from the SRG/eROSITA all-sky survey are shown in fig.~\ref{fig:dir_image}.

When processing the spectroscopic images, we extracted the spectra and measured the redshifts for all of the galaxies that fell on the spectrograph slit. At most the redshifts of six galaxies from one cluster were measured. The light from all these galaxies fell on a 2\arcsec-wide slit during the BTA observations. In some of the galaxy clusters there is a bight cD galaxy in the central cluster region that is brighter than other brightest red-sequence galaxies approximately by 1~mag. Therefore, for some of the galaxy clusters the redshifts were measured from the cD galaxy. The observations of the galaxy clusters for which the redshifts were measured from one galaxy were carried out mostly at AZT-33IK and RTT-150, where the position of the spectrograph slit was fixed. In some of the cases where the measured redshift differs from its photometric estimate, we carried out additional observations of other bright red-sequence galaxies at AZT-33IK and RTT-150.

\begin{figure*}
  \centering
  SRGe\,J144035.3$+$661630 (RC2500)
  \bigskip

  \smfiguresmall{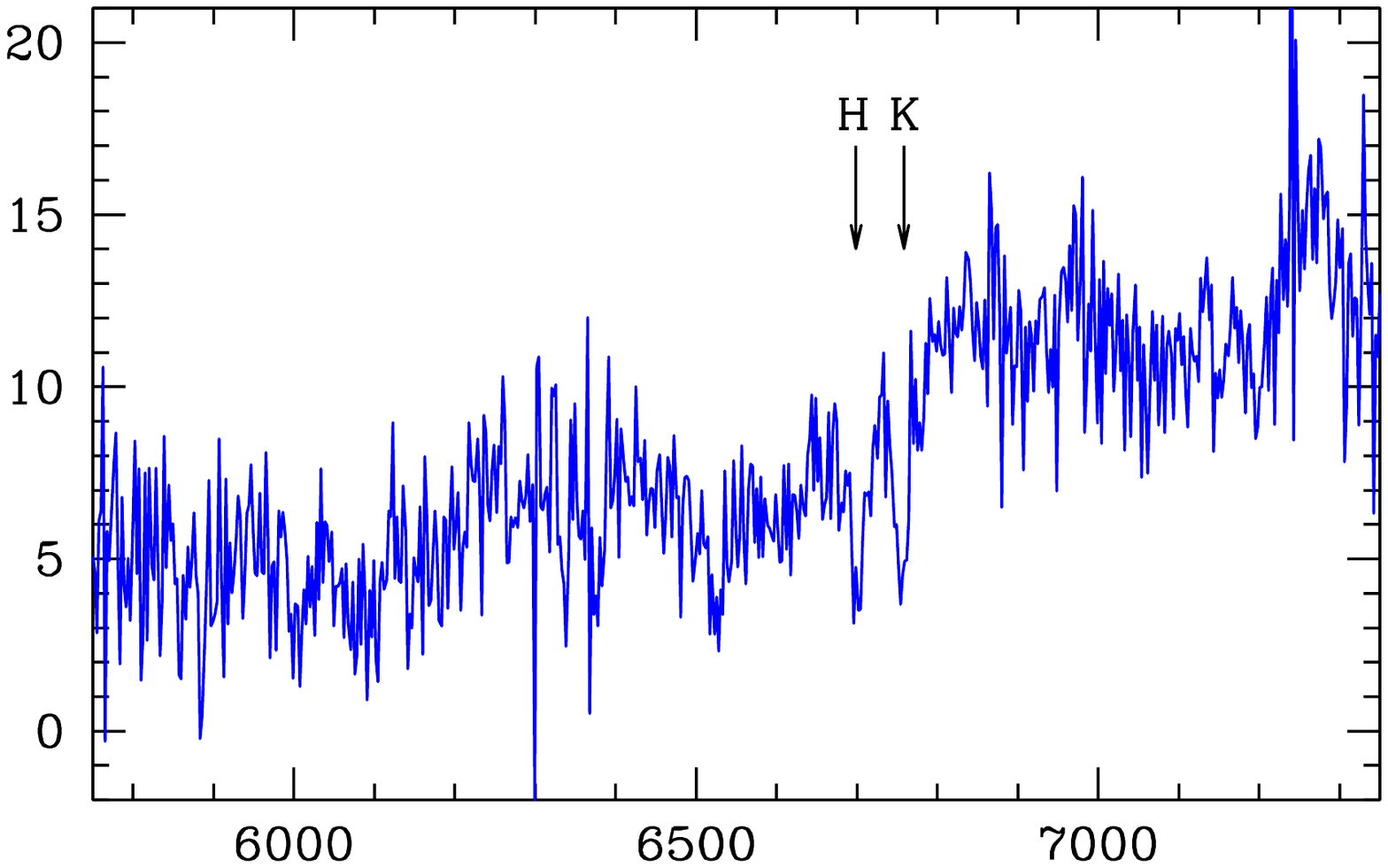}{$\lambda$, \AA}{Flux, $10^{-17}$~erg\,s$^{-1}$\,cm$^{-2}$}~
  \smfiguresmall{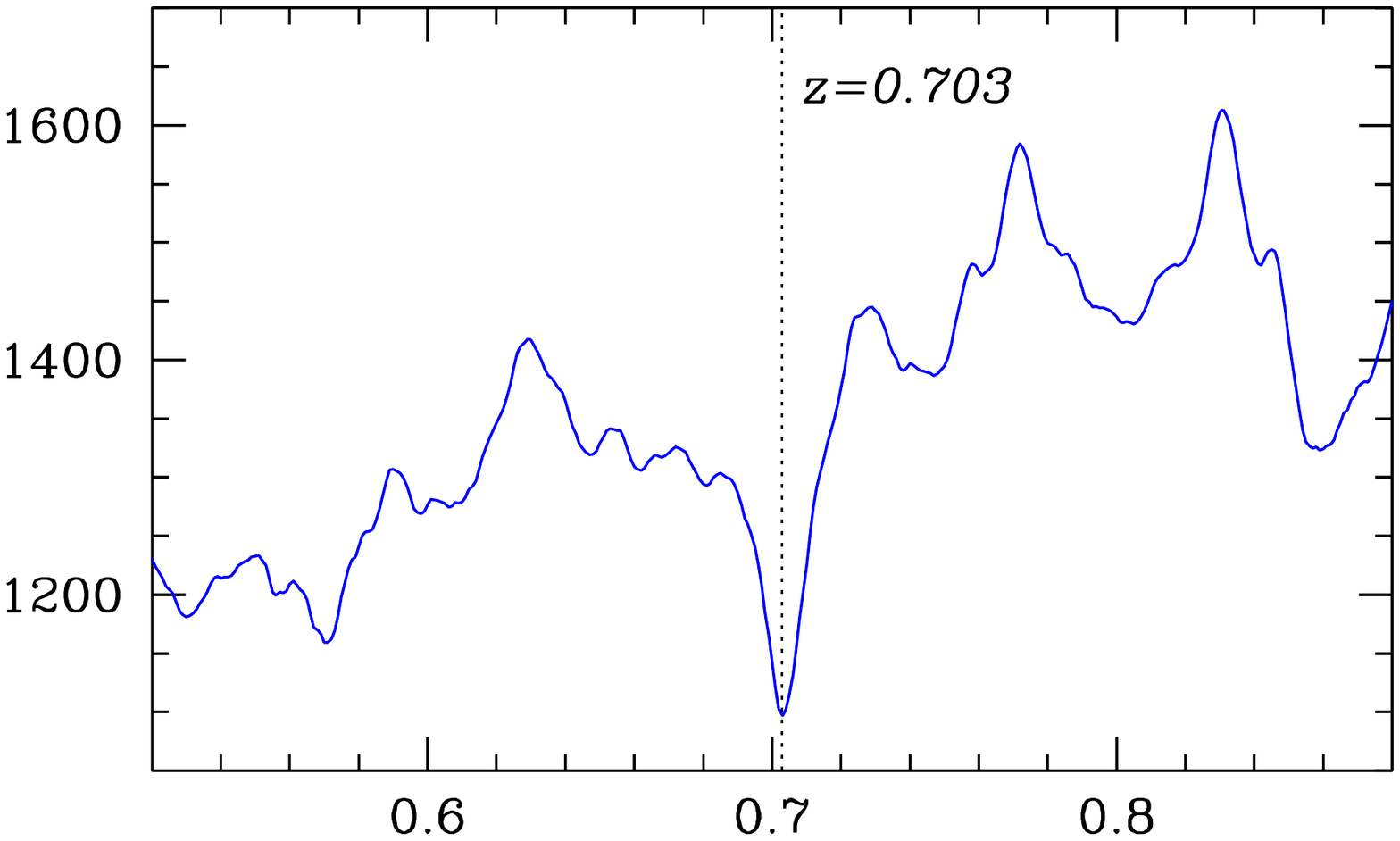}{$z$}{$\chi^2$}
  \bigskip

  \centering
  SRGe\,J172643.2$+$653917 (AZT-33IK)
  \bigskip
  
  \smfiguresmall{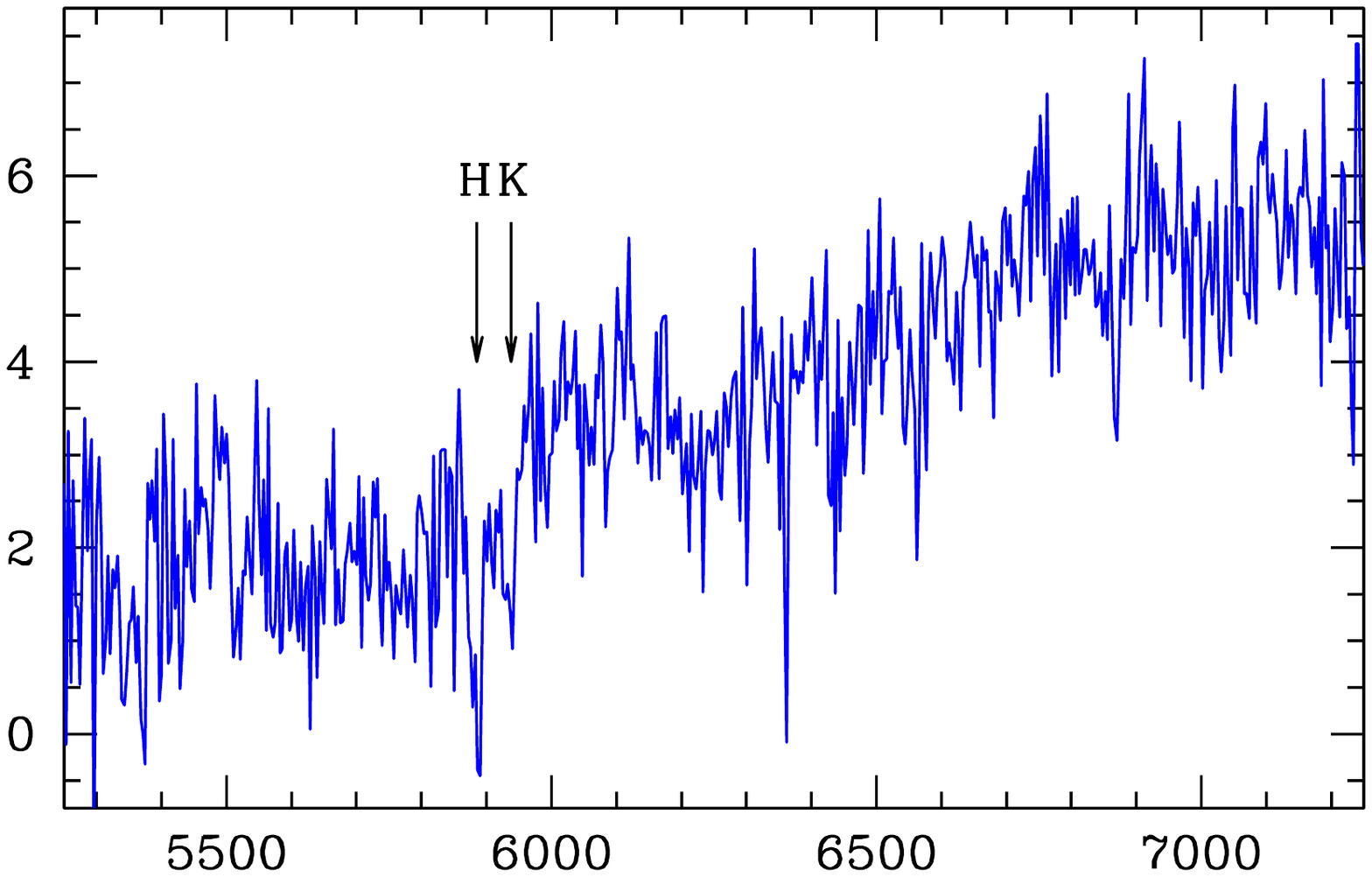}{$\lambda$, \AA}{Flux, $10^{-17}$~erg\,s$^{-1}$\,cm$^{-2}$}~
  \smfiguresmall{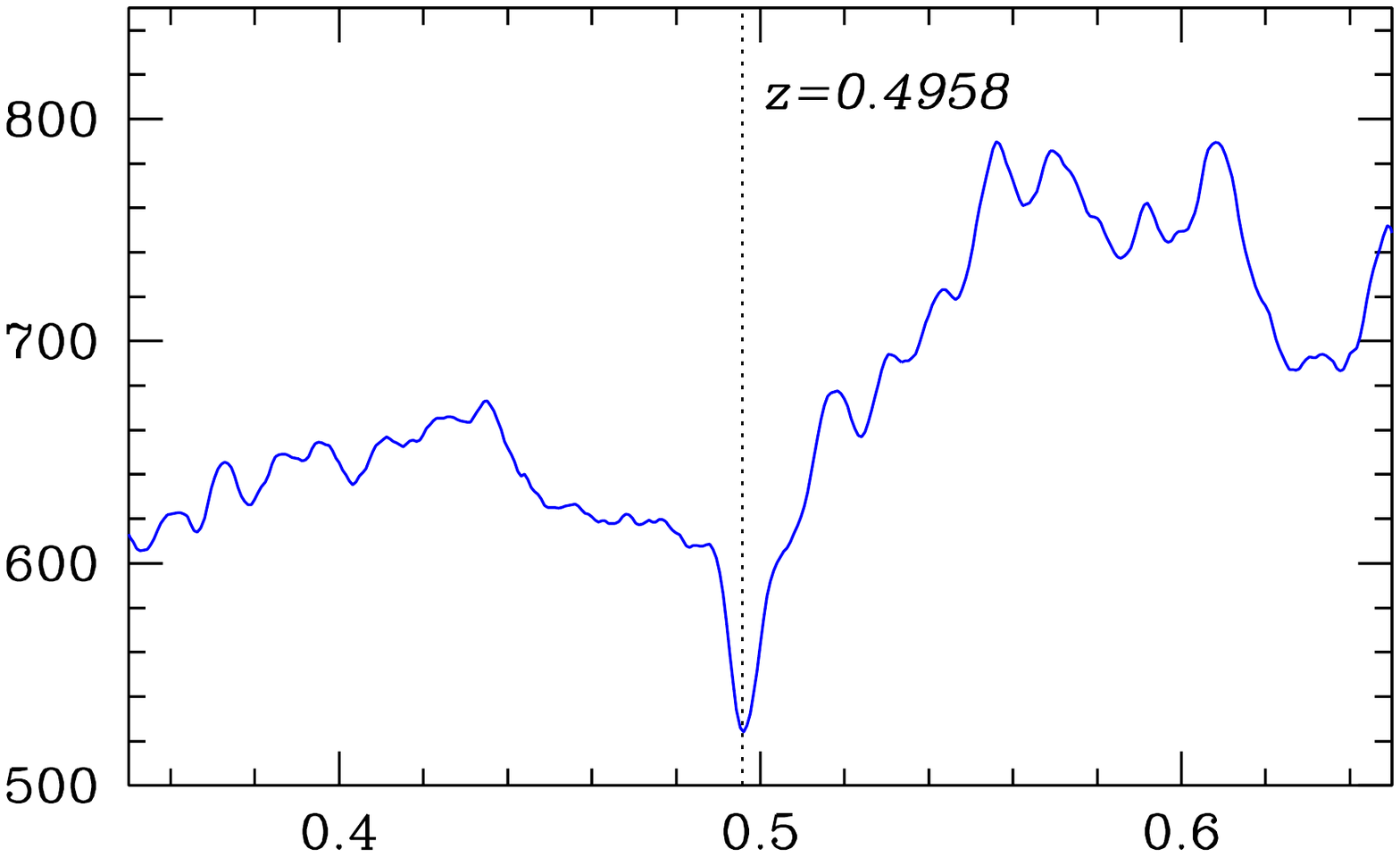}{$z$}{$\chi^2$}
  \bigskip

  \centering
  SRGe\,J065344.7$+$552326 (RTT-150)
  \bigskip

  \smfiguresmall{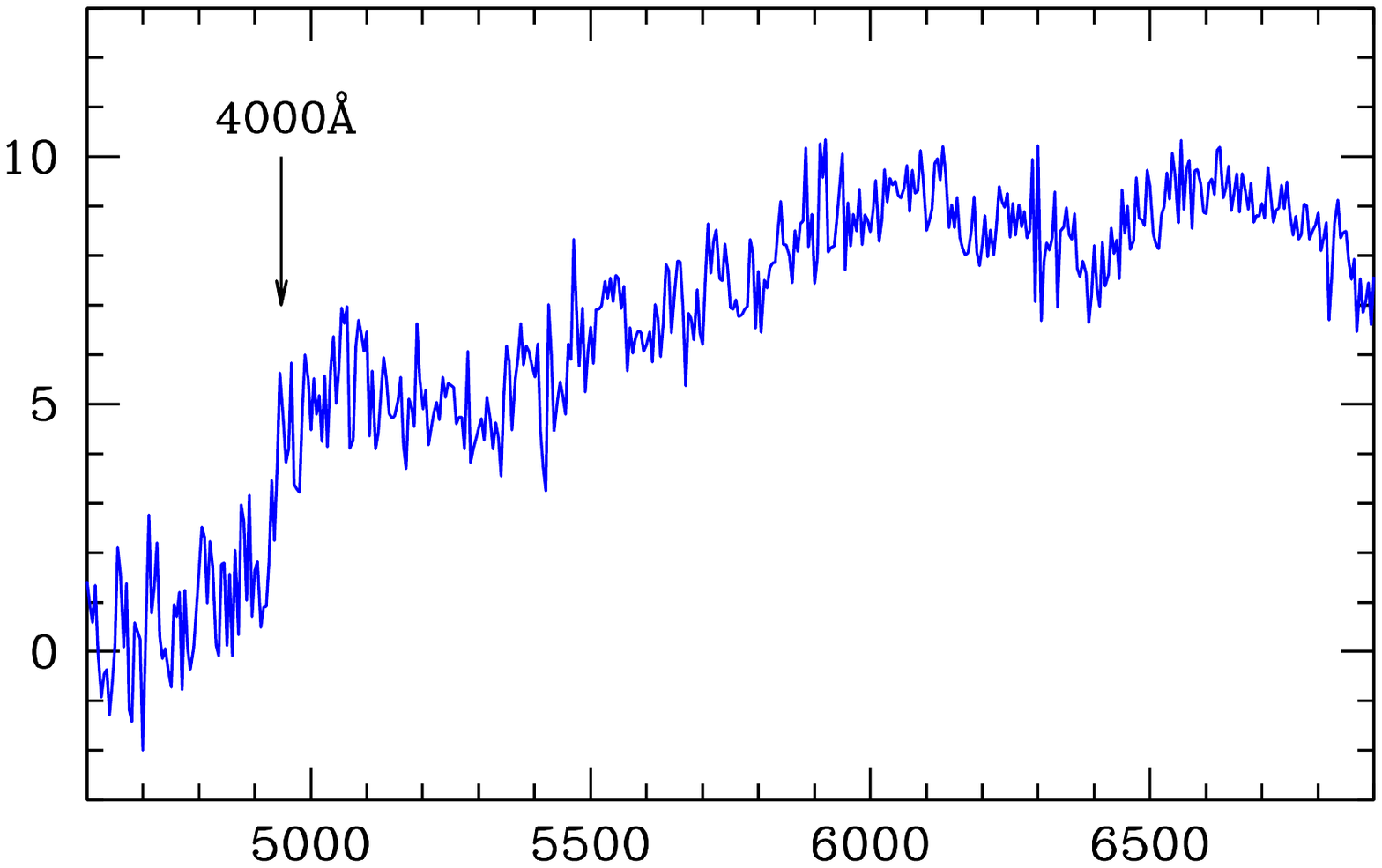}{$\lambda$, \AA}{Flux, $10^{-17}$~erg\,s$^{-1}$\,cm$^{-2}$}~
  \smfiguresmall{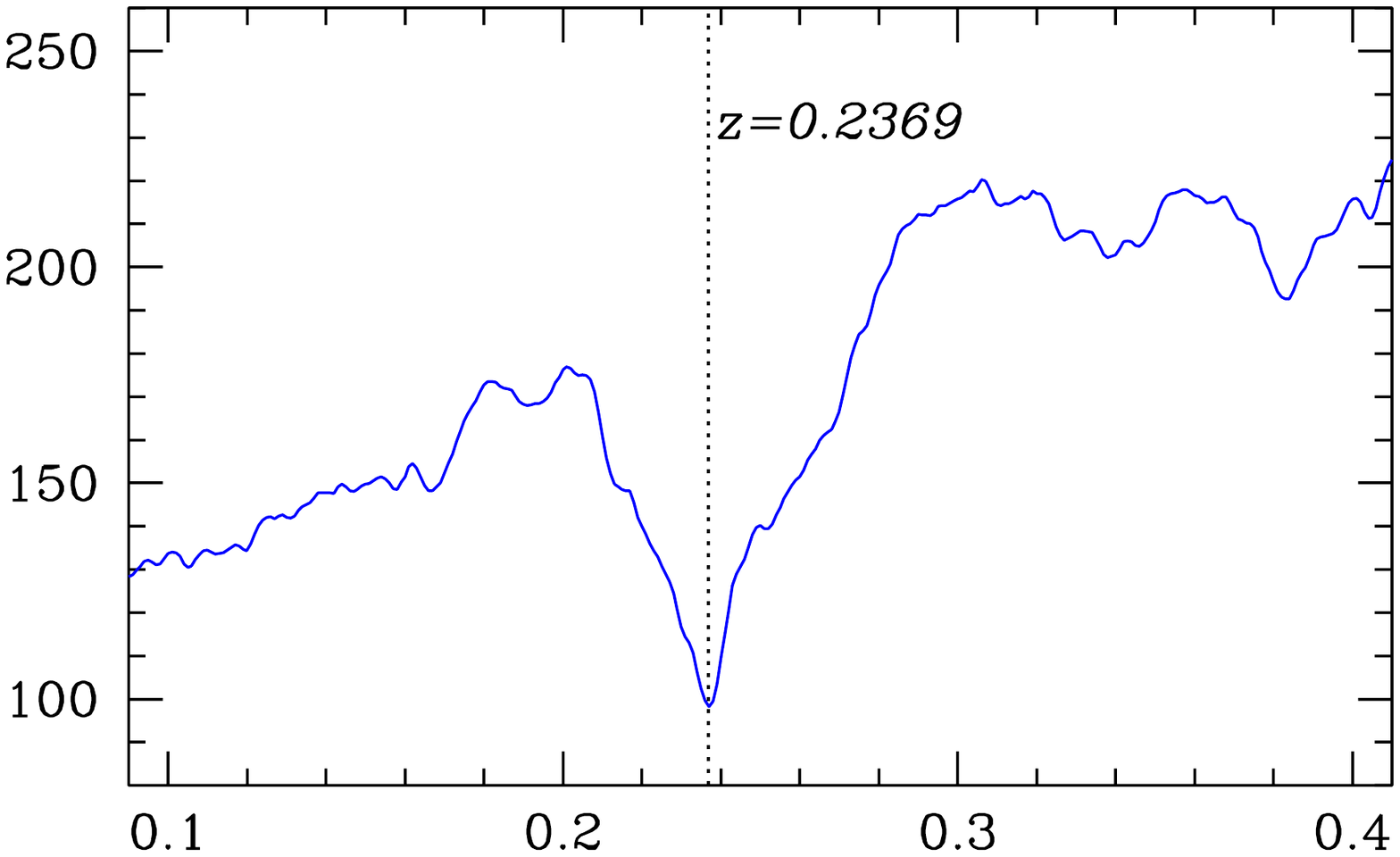}{$z$}{$\chi^2$}
  \bigskip

  \caption{Examples of the spectroscopic redshift measurements for clusters obtained at the telescopes: the 2.5-m CMO telescope (upper row), AZT-33IK (middle row), and RTT-150 (lower row). Left: the spectrum of the brightest cluster galaxy with an indication of some spectral features. Right: the $\chi^2$ value obtained by comparing this spectrum with the template spectrum of an elliptical galaxy.
  }
  \label{fig:spec:example}
\end{figure*}

\begin{figure*}
  \centering
  \hspace{0.8cm}SRGe\,J191751.9$+$692812 ($z_{spec} = 1.096$) \hspace{2.2cm}SRGe\,J215157.4$+$111248 ($z_{spec} = 0.925$)
  \bigskip

  \smfiguresmall{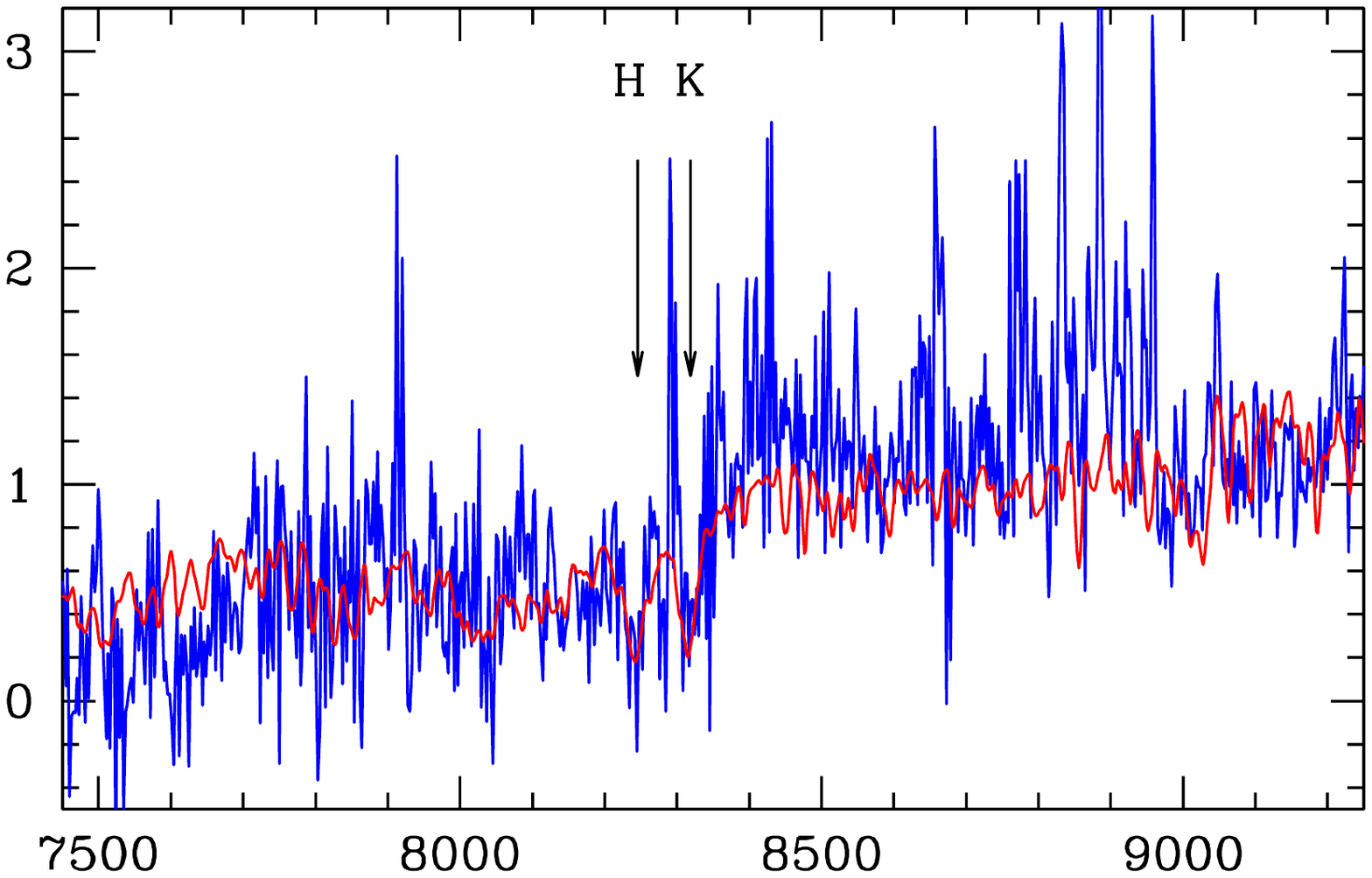}{$\lambda$, \AA}{Flux, $10^{-17}$~erg\,s$^{-1}$\,cm$^{-2}$}~
  \smfiguresmall{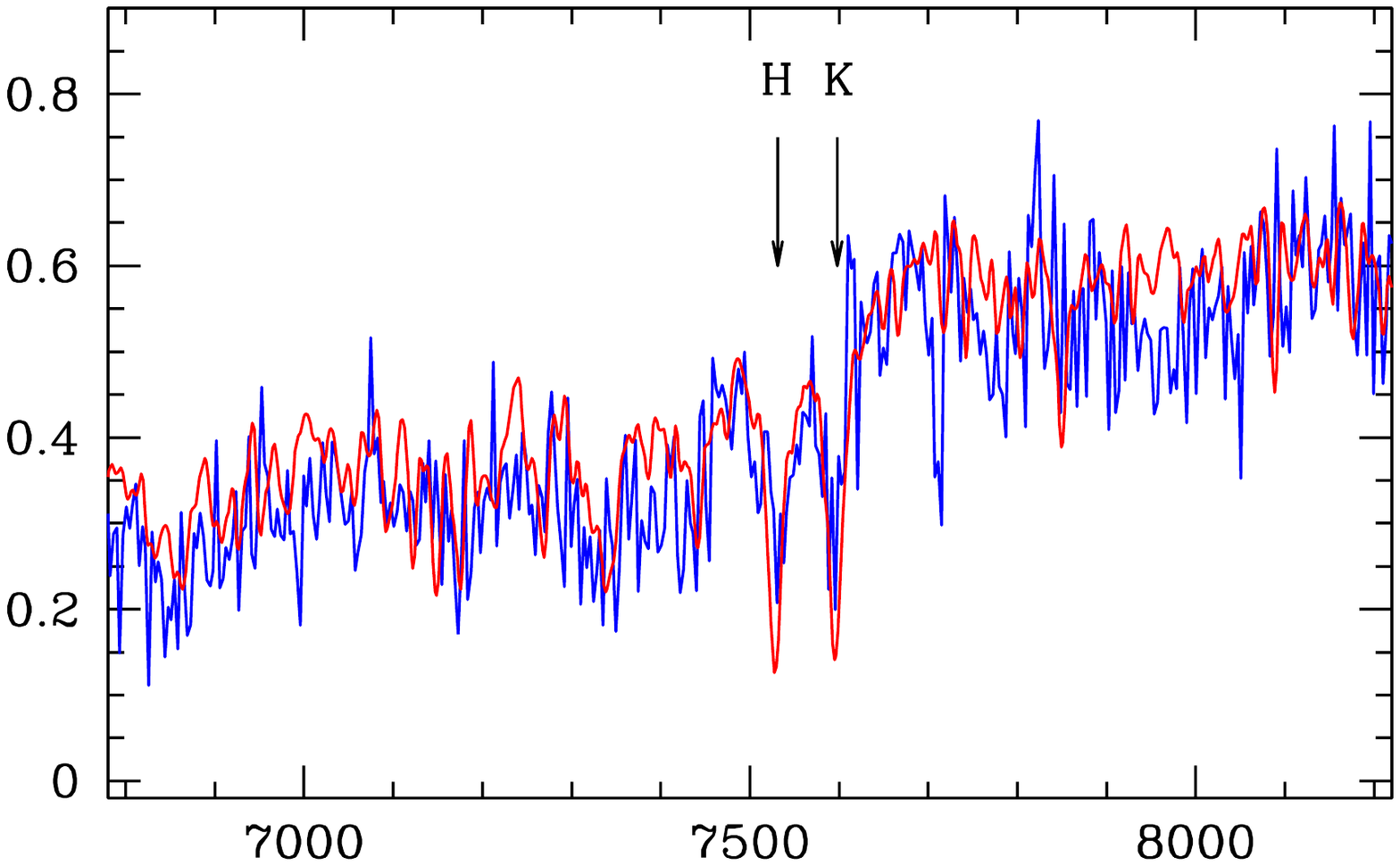}{$\lambda$, \AA}{Flux, $10^{-17}$~erg\,s$^{-1}$\,cm$^{-2}$}
  \bigskip

  \centering
  \hspace{0.8cm}SRGe\,J191842.1$+$744327 ($z_{spec} = 1.024$) \hspace{2.2cm}SRGe\,J132950.1$+$564752 ($z_{spec} = 1.298$)
  \bigskip
  
  \smfiguresmall{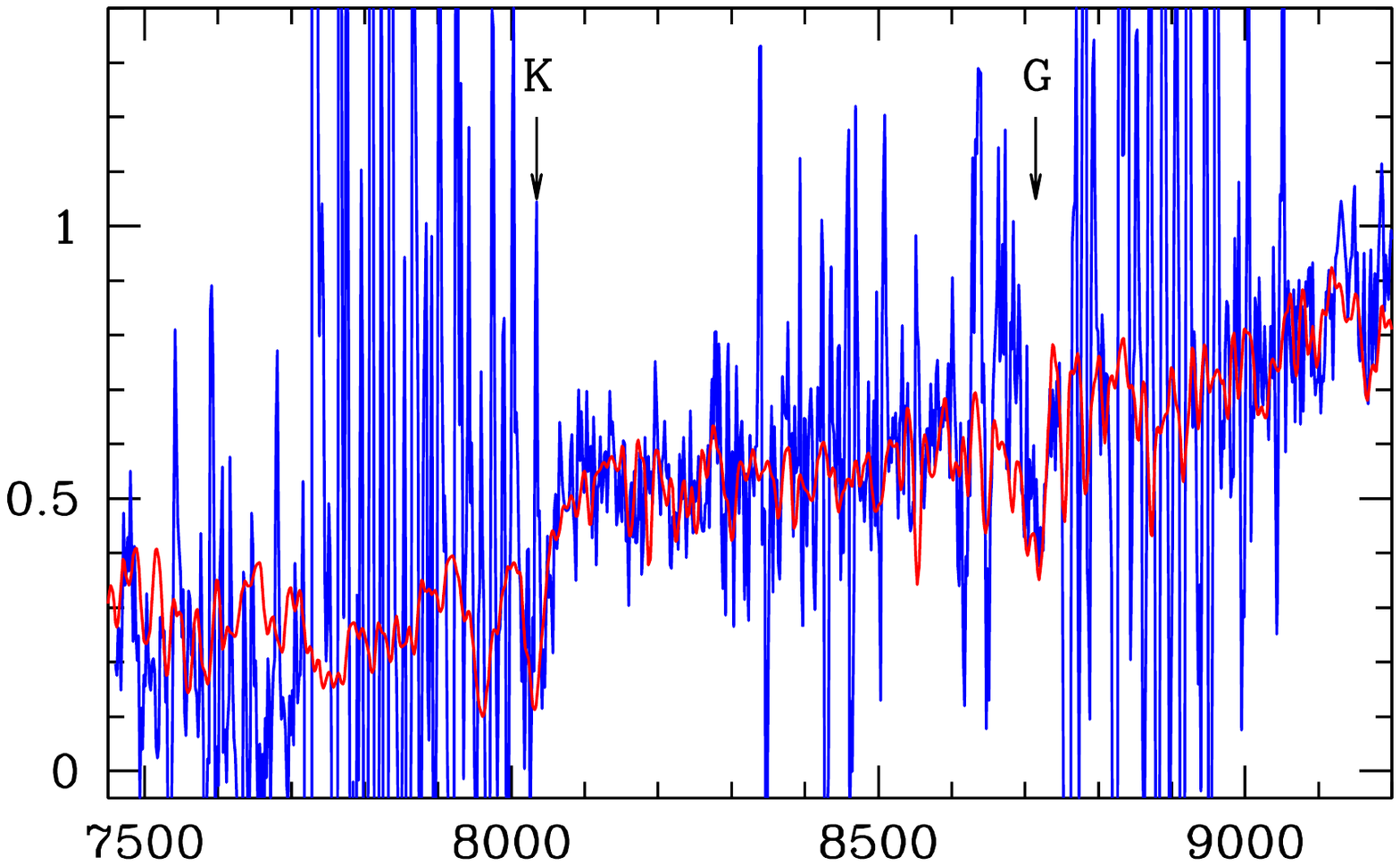}{$\lambda$, \AA}{Flux, $10^{-17}$~erg\,s$^{-1}$\,cm$^{-2}$}~
  \smfiguresmall{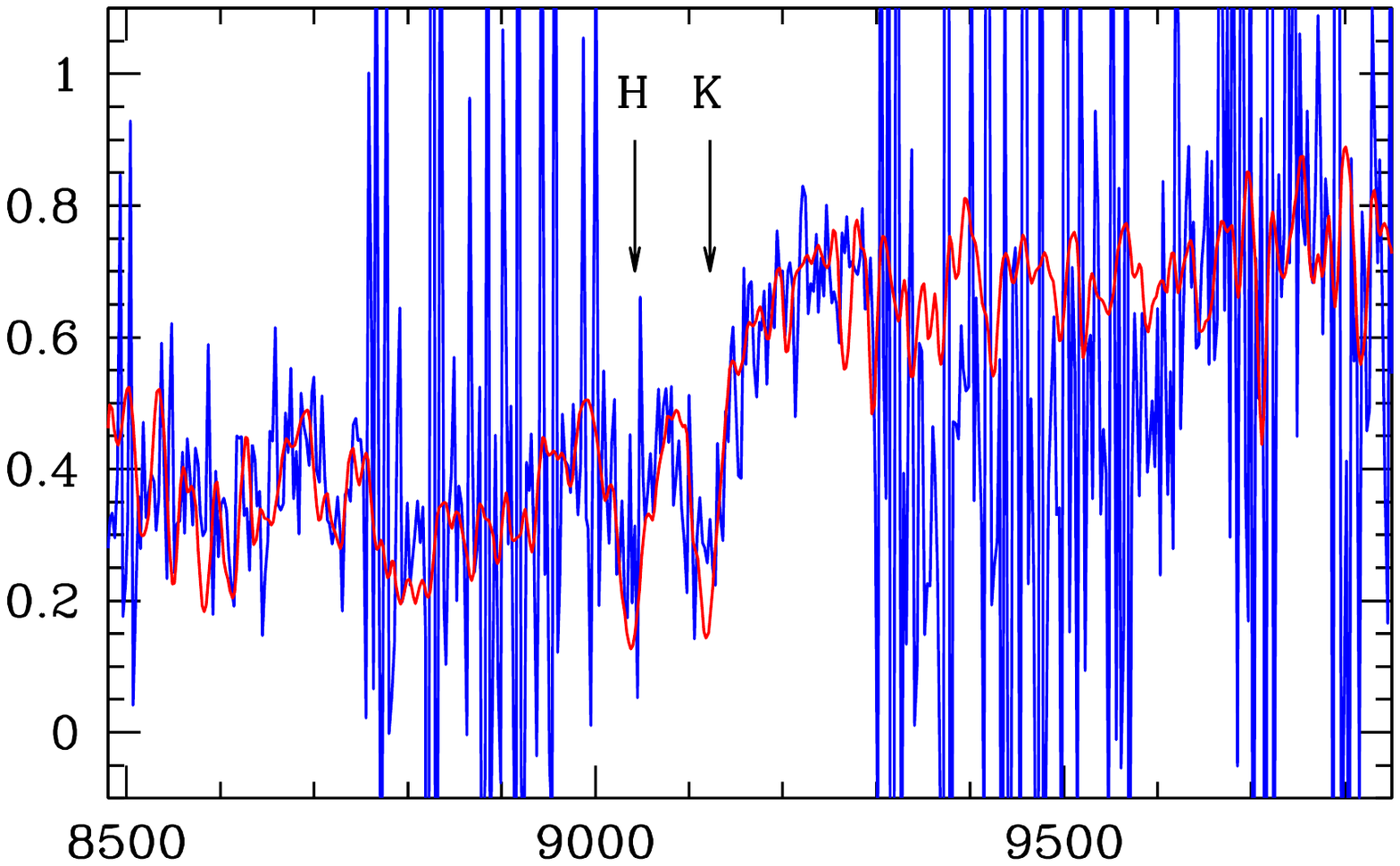}{$\lambda$, \AA}{Flux, $10^{-17}$~erg\,s$^{-1}$\,cm$^{-2}$}
  \bigskip

  \caption{Examples of the spectroscopic redshift measurements for the brightest galaxies of the most distant clusters detected in the SRG/eROSITA all-sky survey that have not been published previously. The blue line indicates the galaxy spectra taken at BTA with the SCORPIO-2 spectrograph, the red line indicates the template comparison spectra. The sum of the spectra of two cluster galaxies is shown for the cluster SRGe\,J191751.9$+$692812.
  }
  \label{fig:bta:hz}
\end{figure*}

The redshifts were measured by comparing the spectra of galaxies with the spectrum of a synthetic stellar population (template) with an age of 11~Gyr and metallicity $Z = 0.02$. For a few most distant galaxy clusters the redshifts were measured using templates with different ages and metallicities. The redshifts of galaxies were determined as the local minimum of the $\chi^2$-distribution obtained by comparing the spectra with the template spectra. If there is a cD galaxy in clusters, then the redshifts of such galaxy clusters were determined as the redshift of the cD galaxy. In the absence of a cD galaxy, the redshifts of galaxy clusters were determined as the arithmetic mean of the spectroscopic redshifts of the brightest galaxies in the central cluster regions. Examples of the spectra for galaxies from the all-sky survey are shown in fig.~\ref{fig:spec:example}. The left and right panels present the spectra of galaxies and their $\chi^2$ distributions obtained by comparing the spectra with the template, respectively. The BTA spectra of the most distant galaxy clusters from the all-sky survey are presented in comparison with the template spectra in fig.~\ref{fig:bta:hz}.

\subsection{Results}

The results of our redshift measurements for the galaxy clusters are presented in Table~\ref{tab:zspec}. The first column gives the eROSITA sources names. The equatorial coordinates of the centers of the X-ray sources ($\alpha$, $\delta$) at the epoch J2000.0 are given in their names. The second and third columns provide the redshifts of the galaxy clusters and the number of galaxies from the clusters whose redshifts were measured, respectively. The accuracy of the redshift measurements for each galaxy cluster is different and, therefore, different numbers of significant figures are given for different galaxy clusters. Note that we took into account the redshifts of only the brightest galaxies closest to the galaxy cluster center. The fourth and fifth columns give the names of the telescopes at which the observations were carried out and the total exposure time of our spectroscopic images. In the case of galaxy cluster observations at several telescopes, the exposure times were added. The next-to-last column gives galaxy clusters catalogs in which they are present. The last column gives notes.

The notes in Table~\ref{tab:zspec} provide information about the previously confirmed or possible presence of weak lensing of galaxies by clusters (lens). The names of the catalogs in which the spectroscopic redshift measurements for the galaxies that belong to the cluster but are not the most massive members of the red sequences are given are also specified. Some of the most massive cluster galaxies have narrow emission lines whose flux ratio is typical for star-forming galaxies. The galaxy clusters in which one or more massive galaxies at the cluster center have a bright [OII]$\lambda$3727\,\AA\ emission line are specified. For several such clusters below we prepare the notes with an estimate of the star formation rate in their cD galaxies from the [OII]$\lambda$3727\,\AA\ line emission.

Some of the galaxy clusters from Table~\ref{tab:zspec} have already been published previously in \cite{br21} (B21) and \cite{br22} (B22). These galaxy clusters also enter into the program of observations of galaxy clusters from the SRG/eROSITA all-sky survey. Four galaxy clusters have been observed previously in medium-band filters with the Hubble Space Telescope (HST). These papers provide only the photometric redshift estimates based on HST observations. Therefore, for these galaxy clusters we obtained spectroscopic redshift measurements. The galaxy clusters being discussed in the text below are marked by an asterisk.

\begin{table*}
  \caption{Galaxy clusters from the all-sky survey}
  \label{tab:zspec}
  \vskip 5mm
  \renewcommand{\arraystretch}{1.0}
  \renewcommand{\tabcolsep}{0.15cm}
  \centering
  \footnotesize
  \begin{tabular}{clccclc}
    \noalign{\doubleline}
    Source name & $z_{spec}$ & $N_{gal}$ & Telescope & Time, min & Catalogs & Notes\\
    \noalign{\vskip 3pt\hrule\vskip 5pt}
    SRGe\,J001104.8$+$272243 & 0.3212 & 3 & BTA & 60 & WHL &  \\
    SRGe\,J001321.8$-$150316 & 0.458 & 2 & RC2500 & 80 & ACT &  \\
    SRGe\,J001502.8$-$151603 & 0.3353 & 2 & BTA & 45 & ACT &  \\
    SRGe\,J001525.5$-$173051 & 0.4622 & 1 & AZT-33IK & 40 & PSZ1,ACT &  \\
    SRGe\,J001735.6$-$114916 & 0.500 & 2 & RC2500 & 120 & - &  \\
    SRGe\,J002641.7$-$152613 & 0.5946 & 3 & BTA & 80 & ACT,WHL &  \\
    SRGe\,J003542.0$+$125554 & 0.247 & 2 & RC2500 & 60 & RM,NSC,NSCS,WHL &  \\
    SRGe\,J004431.2$-$175026 & 0.331 & 2 & RTT-150 & 120 & - &  \\
    SRGe\,J010345.3$+$011836 & 0.199 & 2 & RC2500 & 40 & - &  \\
    SRGe\,J011738.7$-$131454 & 0.1052 & 1 & RC2500 & 20 & - &  \\
    SRGe\,J011817.8$-$215034 & 0.4727 & 2 & BTA & 50 & - &  \\
    SRGe\,J012531.7$+$192251 & 0.5811 & 2 & BTA & 60 & ACT & SDSS\\
    SRGe\,J012711.9$-$162606 & 0.2509 & 2 & AZT-33IK & 40 & ACT &  \\
    SRGe\,J012915.6$-$134233 & 0.376 & 2 & RC2500 & 80 & ACT &  \\
    SRGe\,J013303.5$+$403457 & 0.3864 & 3 & BTA & 45 & - &  \\
    SRGe\,J013923.7$-$112235 & 0.277 & 5 & RC2500 & 80 & PSZ2,ACT &  \\
    SRGe\,J014500.2$+$304113 & 0.7480 & 3 & BTA & 180 & - &  \\
    SRGe\,J015010.0$+$214332 & 0.5658 & 2 & BTA & 45 & - & SDSS\\
    SRGe\,J015448.2$+$211639 & 0.593 & 1 & RC2500 & 100 & - &  \\
    SRGe\,J015747.8$+$381210 & 0.6827 & 4 & BTA & 60 & - & B22\\
    SRGe\,J020037.1$+$373836 & 0.3139 & 4 & BTA & 45 & - &  \\
    SRGe\,J020244.9$+$384733 & 0.3068 & 3 & BTA & 30 & - &  \\
    SRGe\,J020629.8$+$363229 & 0.5164 & 3 & BTA & 40 & - & B22\\
    SRGe\,J021228.1$+$112737 & 0.3990 & 2 & BTA & 45 & - &  \\
    SRGe\,J021250.3$-$112414 & 0.3275 & 1 & AZT-33IK & 40 & ACT &  \\
    SRGe\,J021821.9$+$323337 & 0.4473 & 1 & RC2500 & 120 & WHL & OII3727\\
    SRGe\,J022017.6$-$102414 & 0.4572 & 3 & BTA & 60 & ACT,WHL &  \\
    SRGe\,J022418.2$+$301934 & 0.480 & 2 & RC2500 & 100 & - & lens? *\\
    SRGe\,J023005.7$+$194915 & 0.3368 & 3 & BTA & 60 & - &  \\
    SRGe\,J023203.5$+$222043 & 0.305 & 1 & RTT-150 & 90 & WHL &  \\
    SRGe\,J023251.9$+$102615 & 0.4510 & 1 & RTT-150 & 150 & ACT &  \\
    SRGe\,J023820.8$+$200556 & 0.4205 & 2 & BTA & 30 & WHL & B22\\
    SRGe\,J024102.2$+$255716 & 0.5732 & 4 & BTA & 105 & - & B22, OII3727\\
    SRGe\,J024108.6$+$141044 & 0.5539 & 5 & BTA & 60 & ACT &  \\
    SRGe\,J024140.2$+$220924 & 0.438 & 2 & RC2500, AZT-33IK & 60 & - &  \\
    SRGe\,J024212.7$+$054452 & 0.2871 & 1 & RC2500, AZT-33IK & 120 & WHL,NSC &  \\
    SRGe\,J024450.4$+$264715 & 0.5176 & 2 & BTA & 60 & - &  \\
    SRGe\,J024505.8$+$255725 & 0.3339 & 2 & RC2500, AZT-33IK & 80 & - &  \\
    SRGe\,J024508.5$+$184217 & 0.4405 & 1 & AZT-33IK & 30 & ACT & OII3727\\
    SRGe\,J024614.1$+$103920 & 0.4209 & 1 & AZT-33IK & 40 & - &  \\
    SRGe\,J024716.9$+$335442 & 0.2339 & 1 & BTA & 50 & WHL &  \\
    SRGe\,J025827.1$+$174157 & 0.3859 & 3 & BTA & 75 & - &  \\
    SRGe\,J030255.1$+$841349 & 0.4436 & 2 & BTA & 60 & - &  \\  
    SRGe\,J030526.0$+$094805 & 0.505 & 2 & RC2500 & 180 & ACT,WHL &  \\ 
    SRGe\,J030926.9$+$041339 & 0.458 & 1 & RC2500 & 320 & ACT &  \\
    SRGe\,J032452.5$+$843310 & 0.4138 & 1 & BTA, AZT-33IK & 45 & - &  \\
    SRGe\,J032444.4$+$113942 & 0.4794 & 2 & BTA & 60 & - &  \\
    SRGe\,J033748.0$+$074743 & 0.4733 & 2 & BTA & 60 & WHL &  \\
    SRGe\,J033950.5$+$280522 & 0.1755 & 3 & BTA & 40 & - &  \\
    SRGe\,J034028.9$+$071047 & 0.2336 & 2 & AZT-33IK & 40 & WHL &  \\
    SRGe\,J034821.9$+$205702 & 0.2607 & 2 & BTA & 70 & - &  \\
    SRGe\,J034900.1$+$213302 & 0.3478 & 3 & BTA & 60 & - & B22\\
    SRGe\,J034934.9$+$110529 & 0.268 & 2 & RC2500 & 80 & WHL &  \\
    SRGe\,J035114.1$+$174311 & 0.1565 & 3 & BTA & 30 & WHL &  \\
    \noalign{\vskip 3pt\hrule\vskip 5pt}
    \label{tab:zspec}
    \end{tabular}
\end{table*}

\setcounter{table}{0}
\begin{table*}
  \caption{Contd.}
  \vskip 5mm
  \renewcommand{\arraystretch}{1.0}
  \renewcommand{\tabcolsep}{0.15cm}
  \centering
  \footnotesize
  \begin{tabular}{clccclc}
    \noalign{\doubleline}
    Source name & $z_{spec}$ & $N_{gal}$ & Telescope & Time, min & Catalogs & Notes\\
    \noalign{\vskip 3pt\hrule\vskip 5pt}
    SRGe\,J035322.3$+$110340 & 0.2689 & 3 & BTA & 30 & ACT,WHL,NSC &  \\    
    SRGe\,J040249.5$+$135319 & 0.6194 & 3 & BTA & 60 & ACT &  \\ 
    SRGe\,J040342.7$+$154444 & 0.5300 & 1 & AZT-33IK & 80 & - &  \\
    SRGe\,J040844.2$+$133350 & 0.8013 & 1 & BTA & 180 & - &  \\
    SRGe\,J041402.7$+$830258 & 0.5909 & 1 & BTA & 60 & WHL & B22\\
    SRGe\,J043602.9$+$780337 & 0.5219 & 1 & AZT-33IK & 105 & - &  \\
    SRGe\,J052742.4$+$800205 & 0.3843 & 2 & AZT-33IK & 150 & - &  \\
    SRGe\,J062552.1$+$711624 & 0.7135 & 2 & BTA & 90 & - &  \\
    SRGe\,J063357.4$+$574050 & 0.2361 & 3 & BTA & 50 & - &  \\
    SRGe\,J064609.8$+$764805 & 0.4139 & 2 & BTA & 60 & - &  \\
    SRGe\,J065344.7$+$552326 & 0.2369 & 1 & RTT-150 & 150 & - &  \\
    SRGe\,J065908.8$+$811419 & 0.3614 & 2 & AZT-33IK & 120 & - &  \\
    SRGe\,J070623.4$+$484612 & 0.3777 & 1 & AZT-33IK & 70 & - &  \\
    SRGe\,J071221.3$+$593220 & 0.3315 & 1 & AZT-33IK & 40 & MACS & HST\\
    SRGe\,J072328.4$+$520707 & 0.6303 & 3 & BTA & 120 & - &  \\
    SRGe\,J072935.4$+$643314 & 0.3640 & 2 & AZT-33IK & 90 & - &  \\
    SRGe\,J073222.5$+$551230 & 0.4145 & 2 & AZT-33IK & 90 & - &  \\
    SRGe\,J073609.4$+$515822 & 0.4795 & 1 & AZT-33IK & 70 & - &  \\
    SRGe\,J073723.9$+$544908 & 0.2590 & 1 & AZT-33IK & 80 & - &  \\
    SRGe\,J074554.3$+$800654 & 0.5270 & 3 & BTA & 80 & - &  \\
    SRGe\,J074659.4$+$693740 & 0.5728 & 1 & BTA & 45 & - &  \\
    SRGe\,J075537.4$+$562626 & 0.4527 & 1 & AZT-33IK, RTT-150 & 210 & - &  \\
    SRGe\,J080501.8$+$692700 & 0.5382 & 1 & RTT-150 & 120 & - &  \\
    SRGe\,J080514.7$+$745357 & 0.7491 & 2 & BTA & 70 & - &  \\
    SRGe\,J083241.8$+$665352 & 0.180 & 2 & RC2500 & 60 & WHL,NSC &  \\
    SRGe\,J085156.5$+$744249 & 0.4195 & 1 & AZT-33IK & 75 & - &  \\
    SRGe\,J085213.7$+$783256 & 0.5347 & 2 & RC2500, AZT-33IK & 70 & - &  \\
    SRGe\,J090416.6$+$725951 & 0.2419 & 3 & AZT-33IK & 50 & NSC &  \\
    SRGe\,J090723.8$+$671121 & 0.4911 & 2 & BTA & 45 & - &  \\
    SRGe\,J091600.3$+$754812 & 0.1770 & 2 & BTA & 50 & NSC &  \\
    SRGe\,J100900.9$+$721439 & 0.4441 & 1 & BTA & 20 & - &  \\
    SRGe\,J102534.9$+$755226 & 0.6684 & 1 & RC2500 & 140 & - &  \\
    SRGe\,J103538.2$+$700645 & 0.4383 & 1 & BTA & 30 & - &  \\
    SRGe\,J104446.3$+$723400 & 0.441 & 1 & AZT-33IK & 80 & - &  \\
    SRGe\,J104533.9$+$781319 & 0.3953 & 1 & AZT-33IK & 70 & WHL &  \\
    SRGe\,J105036.9$+$455126 & 0.8747 & 1 & BTA & 120 & - & SDSS\\
    SRGe\,J112840.5$+$763600 & 0.4782 & 1 & AZT-33IK & 80 & WHL &  \\
    SRGe\,J113857.3$+$784117 & 0.6734 & 1 & RC2500 & 160 & - &  \\
    SRGe\,J115342.3$+$771302 & 0.3782 & 2 & BTA, AZT-33IK & 30 & - &  \\
    SRGe\,J115858.1$+$705437 & 0.6337 & 1 & BTA & 90 & WHL &  \\
    SRGe\,J120822.6$+$742216 & 0.3846 & 1 & AZT-33IK & 40 & - &  \\
    SRGe\,J122502.7$+$862739 & 0.1972 & 2 & AZT-33IK & 70 & - &  \\
    SRGe\,J123141.4$+$723821 & 0.328 & 1 & RTT-150 & 90 & - &  \\
    SRGe\,J123221.8$+$592400 & 0.675 & 2 & RC2500 & 240 & - & OII3727, SDSS\\
    SRGe\,J124909.0$+$812316 & 0.5373 & 2 & BTA & 60 & - &  \\
    SRGe\,J125053.7$+$862515 & 0.3691 & 1 & AZT-33IK & 60 & - &  \\
    SRGe\,J125121.5$+$313125 & 0.5053 & 2 & RC2500 & 70 & WHL,NSCS & SDSS\\
    SRGe\,J125445.4$+$470151 & 0.3214 & 2 & AZT-33IK & 90 & NSC,NSCS,1RXS &  \\
    SRGe\,J131229.6$+$725048 & 0.5735 & 2 & BTA & 80 & - & lens, *\\
    SRGe\,J131253.5$+$725502 & 0.2955 & 3 & BTA & 50 & - &  \\
    SRGe\,J132420.4$+$691724 & 0.3566 & 1 & AZT-33IK & 30 & - &  \\
    SRGe\,J132810.3$+$524321 & 0.3217 & 2 & AZT-33IK, RTT-150 & 150 & MACS,WHL & SDSS\\    
    SRGe\,J132950.1$+$564752 & 1.298 & 1 & BTA & 240 & MOO & *\\
    SRGe\,J134330.9$+$792821 & 0.4476 & 4 & BTA & 40 & - &  \\
    SRGe\,J135353.3$+$733157 & 0.4780 & 1 & BTA & 60 & - & OII3727, *\\
    \noalign{\vskip 3pt\hrule\vskip 5pt}
    \label{tab:zspec}
    \end{tabular}
\end{table*}

\setcounter{table}{0}
\begin{table*}
  \caption{Contd.}
  \vskip 5mm
  \renewcommand{\arraystretch}{1.0}
  \renewcommand{\tabcolsep}{0.15cm}
  \centering
  \footnotesize
  \begin{tabular}{clccclc}
    \noalign{\doubleline}
    Source name & $z_{spec}$ & $N_{gal}$ & Telescope & Time, min & Catalogs & Notes\\
    \noalign{\vskip 3pt\hrule\vskip 5pt}
    SRGe\,J135628.1$+$793836 & 0.4530 & 2 & AZT-33IK & 60 & - &  \\
    SRGe\,J135900.1$+$672547 & 0.4544 & 3 & AZT-33IK & 150 & - &  \\
    SRGe\,J135917.9$+$744637 & 0.196 & 1 & RC2500 & 100 & NSC &  \\
    SRGe\,J142452.1$+$662238 & 0.2409 & 2 & AZT-33IK, RTT-150 & 70 & NSC &  \\
    SRGe\,J143202.2$+$851737 & 0.5684 & 2 & BTA & 90 & - & QSO *\\
    SRGe\,J144035.3$+$661630 & 0.699 & 2 & RC2500 & 80 & - &  \\
    SRGe\,J144245.4$+$585306 & 0.6238 & 1 & BTA & 90 & - & SDSS\\
    SRGe\,J144605.0$+$753727 & 0.1737 & 2 & AZT-33IK & 50 & - &  \\
    SRGe\,J145131.8$+$810640 & 0.546 & 2 & RC2500 & 120 & - &  \\
    SRGe\,J150225.3$+$653951 & 0.6502 & 6 & BTA & 90 & - & OII3727\\
    SRGe\,J150743.4$+$700724 & 0.5154 & 2 & AZT-33IK & 100 & - &  \\
    SRGe\,J150823.5$+$645304 & 0.2520 & 2 & AZT-33IK & 80 & NSC &  \\
    SRGe\,J151051.9$+$670628 & 0.4196 & 2 & AZT-33IK & 30 & XMM,WHL,NSC &  \\
    SRGe\,J152236.4$+$640543 & 0.702 & 1 & RC2500 & 100 & WHL &  \\
    SRGe\,J152736.6$+$095513 & 0.7574 & 2 & BTA & 60 & ACT &  \\
    SRGe\,J152854.9$+$852009 & 0.2520 & 3 & AZT-33IK & 70 & - &  \\
    SRGe\,J153256.6$+$670715 & 0.7246 & 2 & BTA & 120 & - &  \\
    SRGe\,J154729.2$+$701838 & 0.554 & 1 & RC2500 & 40 & - &  \\
    SRGe\,J155449.4$+$535841 & 0.7434 & 2 & BTA & 90 & - &  \\
    SRGe\,J160002.5$-$035434 & 0.2729 & 1 & RTT-150 & 100 & - &  \\
    SRGe\,J160121.3$+$712533 & 0.3349 & 2 & AZT-33IK, RTT-150 & 40 & - &  \\
    SRGe\,J160205.7$+$650555 & 0.2512 & 2 & RTT-150 & 85 & NSC &  \\
    SRGe\,J160948.4$-$040052 & 0.3495 & 1 & AZT-33IK & 30 & - &  \\
    SRGe\,J161215.6$+$662009 & 0.616 & 1 & RC2500 & 120 & - &  \\
    SRGe\,J161519.8$-$035343 & 0.262 & 3 & RC2500 & 80 & ACT,WHL &  \\
    SRGe\,J162051.3$+$651237 & 0.1685 & 2 & AZT-33IK & 40 & NSC &  \\
    SRGe\,J162125.2$+$723233 & 0.5884 & 2 & BTA & 50 & - &  \\
    SRGe\,J162126.4$+$602653 & 0.2822 & 1 & RTT-150 & 30 & NSC &  \\
    SRGe\,J162545.1$+$672930 & 0.8132 & 1 & BTA & 90 & - &  lens, *\\
    SRGe\,J163405.9$+$632020 & 0.5174 & 1 & BTA & 45 & WHL &  \\
    SRGe\,J164110.0$+$100449 & 0.3371 & 1 & RTT-150 & 90 & WHL &  \\
    SRGe\,J164201.1$+$575631 & 0.292 & 2 & RC2500 & 60 & - &  \\
    SRGe\,J164500.8$+$014009 & 0.337 & 3 & RC2500 & 40 & ACT,MACS,WHL & HST\\
    SRGe\,J164727.5$+$044052 & 0.278 & 2 & RC2500 & 20 & ACT,WHL &  \\
    SRGe\,J165012.0$+$650920 & 0.3835 & 2 & AZT-33IK & 10 & - &  \\
    SRGe\,J165151.8$+$081034 & 0.3288 & 1 & AZT-33IK & 40 & - &  \\
    SRGe\,J165201.9$+$800430 & 0.8131 & 2 & BTA & 105 & - &  \\
    SRGe\,J165218.4$+$553455 & 0.324 & 2 & RC2500 & 20 & MACS & HST\\
    SRGe\,J170139.9$+$474516 & 0.3846 & 4 & BTA & 60 & RM,WHL & SDSS\\
    SRGe\,J170519.0$+$850451 & 0.3036 & 2 & BTA & 30 & - &  \\
    SRGe\,J171430.8$+$481644 & 0.253 & 2 & RC2500 & 60 & - &  \\
    SRGe\,J171621.3$+$325646 & 0.2583 & 2 & RC2500 & 40 & RM,WHL & SDSS\\
    SRGe\,J171628.9$+$193457 & 0.1764 & 1 & RTT-150 & 50 & ACT,Zw,NSC &  \\
    SRGe\,J172037.9$+$212851 & 0.3017 & 2 & RC2500 & 40 & - &  \\
    SRGe\,J172321.5$+$235041 & 0.229 & 1 & RTT-150 & 60 & - & *\\
    SRGe\,J172643.2$+$653917 & 0.4958 & 1 & AZT-33IK & 30 & WHL &  \\
    SRGe\,J173225.2$+$193340 & 0.540 & 2 & RC2500 & 80 & ACT &  \\
    SRGe\,J173349.5$+$330419 & 0.401 & 1 & RTT-150 & 100 & RM,WHL &  \\
    SRGe\,J173355.8$+$300018 & 0.201 & 1 & RTT-150 & 60 & WHL &  \\
    SRGe\,J173721.0$+$334748 & 0.3822 & 1 & RC2500 & 20 & WHL &  \\
    SRGe\,J173807.2$+$600621 & 0.3317 & 2 & BTA & 40 & MACS,WHL & HST\\
    SRGe\,J174014.2$+$594415 & 0.601 & 2 & RC2500 & 60 & WHL &  \\
    SRGe\,J174110.4$+$471720 & 0.4793 & 2 & BTA & 60 & WHL & OII3727\\
    SRGe\,J174334.2$+$420106 & 0.4713 & 1 & BTA & 90 & WHL & \\
    \noalign{\vskip 3pt\hrule\vskip 5pt}
    \label{tab:zspec}
    \end{tabular}
\end{table*}

\setcounter{table}{0}
\begin{table*}
  \caption{Contd.}
  \vskip 5mm
  \renewcommand{\arraystretch}{1.0}
  \renewcommand{\tabcolsep}{0.15cm}
  \centering
  \footnotesize
  \begin{tabular}{clccclc}
    \noalign{\doubleline}
    Source name & $z_{spec}$ & $N_{gal}$ & Telescope & Time, min & Catalogs & Notes\\
    \noalign{\vskip 3pt\hrule\vskip 5pt}
    SRGe\,J174434.9$+$822459 & 0.440 & 1 & RC2500 & 100 & - &  \\
    SRGe\,J174538.1$+$500041 & 0.6285 & 1 & BTA & 60 & WHL & B22\\
    SRGe\,J174629.3$+$472502 & 0.4816 & 6 & BTA & 45 & WHL & lens, *\\
    SRGe\,J175035.3$+$435243 & 0.481 & 1 & RC2500 & 40 & WHL &  \\
    SRGe\,J175052.5$+$415846 & 0.5828 & 2 & BTA & 90 & WHL &  \\
    SRGe\,J175155.8$+$295145 & 0.3582 & 3 & BTA & 60 & - &  \\
    SRGe\,J175307.3$+$583500 & 0.429 & 2 & RC2500 & 80 & WHL &  \\
    SRGe\,J175729.4$+$304539 & 0.6146 & 2 & BTA & 45 & - &  \\
    SRGe\,J180245.8$+$292831 & 0.6816 & 1 & BTA & 150 & - &  \\
    SRGe\,J180303.3$+$374252 & 0.3428 & 2 & BTA & 60 & WHY &  \\
    SRGe\,J181119.7$+$431250 & 0.5183 & 2 & BTA & 70 & WHL &  \\
    SRGe\,J181528.3$+$831620 & 0.5464 & 2 & BTA & 45 & - &  \\
    SRGe\,J182326.3$+$541527 & 0.5335 & 2 & BTA, RTT-150 & 140 & - &  \\
    SRGe\,J182323.1$+$840138 & 0.185 & 2 & RTT-150 & 180 & - &  \\
    SRGe\,J182722.2$+$463850 & 0.4730 & 2 & BTA & 90 & - &  \\    
    SRGe\,J183034.3$+$565339 & 0.7867 & 2 & BTA & 45 & PSZ2 & B22\\
    SRGe\,J183852.4$+$662829 & 0.3951 & 2 & BTA & 50 & - &  \\
    SRGe\,J184121.6$+$535027 & 0.946 & 2 & BTA & 90 & - &  B22\\
    SRGe\,J185131.4$+$653106 & 0.4633 & 2 & BTA & 105 & - &  \\
    SRGe\,J191309.4$+$740148 & 0.2696 & 3 & BTA & 75 & - &  \\
    SRGe\,J191751.9$+$692812 & 1.098 & 2 & BTA & 160 & - &  *\\
    SRGe\,J191842.1$+$744327 & 1.024 & 1 & BTA & 150 & - &  *\\
    SRGe\,J192345.4$+$730903 & 0.2762 & 2 & BTA & 75 & - &  \\
    SRGe\,J192913.6$+$643318 & 0.4295 & 2 & BTA & 60 & - &  \\
    SRGe\,J200210.7$+$780245 & 0.3468 & 4 & BTA & 60 & WHL &  \\
    SRGe\,J204908.2$+$763440 & 0.5085 & 3 & BTA & 90 & - &  \\
    SRGe\,J205158.7$-$013116 & 0.3869 & 4 & BTA & 60 & ACT &  \\
    SRGe\,J210510.1$-$032638 & 0.8374 & 1 & BTA & 50 & - & B22\\    
    SRGe\,J210510.6$-$224911 & 0.7573 & 2 & BTA & 80 & ACT,SPT & B21\\
    SRGe\,J213403.2$+$141635 & 0.3776 & 1 & BTA & 45 & ACT &  \\
    SRGe\,J213701.4$-$130727 & 0.2029 & 2 & RTT-150 & 40 & - &  \\
    SRGe\,J213703.6$-$150529 & 0.2103 & 1 & AZT-33IK & 40 & - & 2dF\\
    SRGe\,J213714.3$-$223216 & 0.6137 & 2 & BTA & 120 & - & B22\\
    SRGe\,J214054.0$-$191035 & 0.5073 & 3 & BTA & 90 & - &  \\
    SRGe\,J215157.4$+$111248 & 0.919 & 2 & BTA & 120 & ACT &  *\\    
    SRGe\,J220744.1$-$041657 & 0.092 & 3 & RTT-150 & 250 & - &  \\
    SRGe\,J221101.3$-$065318 & 0.3697 & 5 & BTA, RTT-150 & 200 & ACT,RM,WHL &  \\
    SRGe\,J221953.2$-$035005 & 0.4306 & 2 & BTA & 45 & ACT,RM,WHL &  \\
    SRGe\,J222556.3$-$123904 & 0.2752 & 1 & RTT-150 & 50 & - & 6dF\\
    SRGe\,J222701.8$-$053047 & 0.2855 & 1 & AZT-33IK & 50 & ACT,RM,WHL &  \\
    SRGe\,J224722.8$+$113354 & 0.6905 & 3 & BTA & 120 & - &  \\
    SRGe\,J224811.8$+$353319 & 0.2238 & 1 & AZT-33IK & 10 & - &  \\
    SRGe\,J225817.2$-$225553 & 0.5279 & 2 & BTA & 60 & ACT,SPT &  \\
    SRGe\,J230019.3$-$133429 & 0.338 & 2 & RC2500 & 40 & - &  \\
    SRGe\,J230237.3$+$353606 & 0.5209 & 2 & BTA & 40 & - &  \\
    SRGe\,J231103.8$-$125846 & 0.405 & 2 & RC2500 & 80 & ACT &  \\
    SRGe\,J231522.4$+$090705 & 0.7350 & 3 & BTA & 120 & ACT & SDSS\\
    SRGe\,J231534.7$-$064746 & 0.3265 & 1 & RTT-150 & 45 & ACT,RM,WHL &  \\
    SRGe\,J231613.5$-$060924 & 0.281 & 3 & RTT-150 & 180 & ACT,RM,WHL &  \\
    SRGe\,J231745.7$-$110412 & 0.7217 & 2 & BTA & 70 & ACT &  \\
    SRGe\,J232006.8$-$120210 & 0.404 & 2 & RC2500 & 60 & ACT &  \\
    SRGe\,J232204.4$+$394150 & 0.6278 & 2 & BTA & 90 & - &  \\
    SRGe\,J234200.4$+$833346 & 0.7551 & 4 & BTA & 60 & - &  \\
    \noalign{\vskip 3pt\hrule\vskip 5pt}
    \end{tabular}
\end{table*}

The redshifts of the galaxies for which the redshifts were measured will be published later at the Strasbourg Astronomical Data Center\footnote{https://cds.u-strasbg.fr/}. The redshifts were measured for a total of $\approx$450 galaxies. An example of the redshifts of galaxies in the field of five galaxy clusters arranged in order of increasing right ascension is given in Table~\ref{tab:gal_zspec}. The first column gives the name of the galaxy cluster field defined by the coordinates of the X-ray source center. The second and third columns give the coordinates of the galaxies. The fourth and fifth columns give the redshifts of the galaxies and, if present, their errors. The redshift measurement accuracies for different galaxies can differ and, therefore, the redshifts are given in the table for different galaxies with different numbers of significant figures. The table also includes the spectroscopic redshift measurements for the background and foreground galaxies the light from which fell on the spectrograph slit by chance during the observations.

\begin{table}
  \caption{Galaxy redshifts. The full version of the table will be published later at the Strasbourg Astronomical Data Center.}
  \label{tab:gal_zspec}
  \vskip 5mm
  \renewcommand{\arraystretch}{1.0}
  \renewcommand{\tabcolsep}{0.08cm}
  \centering
  \footnotesize
  \begin{tabular}{cccll}
    \noalign{\doubleline}
    & \multispan2\hfil Coordinates (J2000)\hfil\\
    Source ﬁeld & $\alpha$ & $\delta$ & $z_{spec}$ & $z_{spec}^{err}$\\
    \noalign{\vskip 3pt\hrule\vskip 5pt}
    SRGe\,J001104.8$+$272243 & 00 11 04.6 & $+$27 22 32 & 0.3193 & 0.0006\\
    SRGe\,J001104.8$+$272243 & 00 11 04.7 & $+$27 22 35 & 0.3217 & 0.0006\\
    SRGe\,J001104.8$+$272243 & 00 11 04.8 & $+$27 22 39 & 0.3228 & 0.0004\\
    SRGe\,J001321.8$-$150316 & 00 13 20.7 & $-$15 02 52 & 0.455 & \\
    SRGe\,J001321.8$-$150316 & 00 13 22.9 & $-$15 03 17 & 0.458 & \\
    SRGe\,J001502.8$-$151603 & 00 15 01.6 & $-$15 16 01 & 0.3335 & 0.0003\\
    SRGe\,J001502.8$-$151603 & 00 15 03.5 & $-$15 16 20 & 0.3370 & 0.0004\\
    SRGe\,J001525.5$-$173051 & 00 15 24.4 & $-$17 30 35 & 0.4622 & 0.0019\\
    SRGe\,J001735.6$-$114916 & 00 17 36.2 & $-$11 49 15 & 0.499 & \\
    SRGe\,J001735.6$-$114916 & 00 17 36.2 & $-$11 49 15 & 0.500 & \\
    SRGe\,J002641.7$-$152613 & 00 26 42.1 & $-$15 26 36 & 0.5929 & 0.0010\\
    SRGe\,J002641.7$-$152613 & 00 26 42.2 & $-$15 26 08 & 0.6016 & 0.0030\\
    SRGe\,J002641.7$-$152613 & 00 26 42.2 & $-$15 26 19 & 0.5963 & 0.0014\\
    & \multispan2\hspace{1.0cm}... & & \\
    \noalign{\vskip 3pt\hrule\vskip 5pt}
    \end{tabular}
\end{table}

\begin{figure}
  \centering
    \includegraphics[width=1\columnwidth]{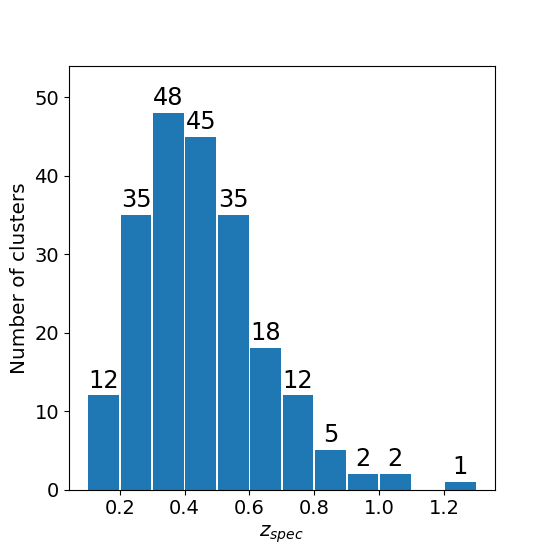}
    \caption{The redshift distribution of the number of galaxy clusters from the SRG/eROSITA survey whose redshifts were measured.
    }
  \label{fig:z-distr}
\end{figure}

The distribution of the number of galaxy clusters for which the redshifts were measured is presented in Fig.~\ref{fig:z-distr}. The distribution does not include one galaxy cluster whose redshift is lower than $z = 0.1$. This distribution shows the current progress of the implementation of our program of spectroscopic measurements of galaxy clusters from the SRG/eROSITA all-sky survey and does not carry the statistical significance.

The galaxy clusters from the SRG/eROSITA all-sky survey were cross-matched with other catalogs of galaxy clusters and catalogs of X-ray sources using NED\footnote{https://ned.ipac.caltech.edu/}. The cross-match radius was selected to be 5\arcmin. Out of the 216 galaxy clusters, we failed to crossmatch 139 clusters with objects from other catalogs. These galaxy clusters have been discovered for the first time. For all of the galaxy clusters in Table~\ref{tab:zspec} that are present in other catalogs we measured their spectroscopic redshifts. Only the photometric redshift estimates are given in other catalogs for these galaxy clusters. Several galaxy clusters have already been known as X-ray sources from the data of the ROSAT \citep{1xrs,macs} and XMM-Newton \citep{xmm} space X-ray observatories. About 40 galaxy clusters are known as Sunyaev–Zeldovich sources from the surveys of the Atacama observatory \citep{act}, the South Pole Telescope \citep{spt}, and the Planck space observatory \citep{psz1,psz2}. In some cases, the redshifts of the cluster galaxies were measured spectroscopically in the SDSS \citep{sdssdr16}, 2df \citep{2dF}, and 6df \citep{2dF} surveys. These galaxies are not among the brightest cluster galaxies and are far from the central regions of the clusters. Therefore, in these cases, it was decided to carry out additional observations to improve the redshifts of these clusters. Table~\ref{tab:cat} gives the lists of catalogs and the number of cross-matched galaxy clusters.

\begin{table}
  \caption{The number of galaxy clusters identified with clusters in other catalogs}
  \label{tab:cat}
  \vskip 5mm
  \renewcommand{\arraystretch}{1.1}
  \renewcommand{\tabcolsep}{0.20cm}
  \centering
  \footnotesize
  \begin{tabular}{lcl}
    \noalign{\doubleline}
    Catalog & Number & Reference\\
    \noalign{\vskip 3pt\hrule\vskip 5pt}
    - & 139 & \\
    WHL & 51 & \cite{whl}\\
    ACT & 37 & \cite{act}\\
    NSC & 15 & \cite{nsc}\\
    RM & 9 & \cite{rm}\\
    MACS & 5 & \cite{macs}\\
    NSCS & 3 & \cite{nscs}\\
    SPT & 2 & \cite{spt}\\
    PSZ2 & 2 & \cite{psz2}\\
    PSZ1 & 1 & \cite{psz1}\\
    Zw & 1 & \cite{zw}\\
    XMM & 1 & \cite{xmm}\\
    WHY & 1 & \cite{why}\\
    MOO & 1 & \cite{moo}\\
    \noalign{\vskip 3pt\hrule\vskip 5pt}
    \label{tab:cat}
    \end{tabular}
\end{table}

\begin{figure}
  \centering
    \includegraphics[width=1\columnwidth]{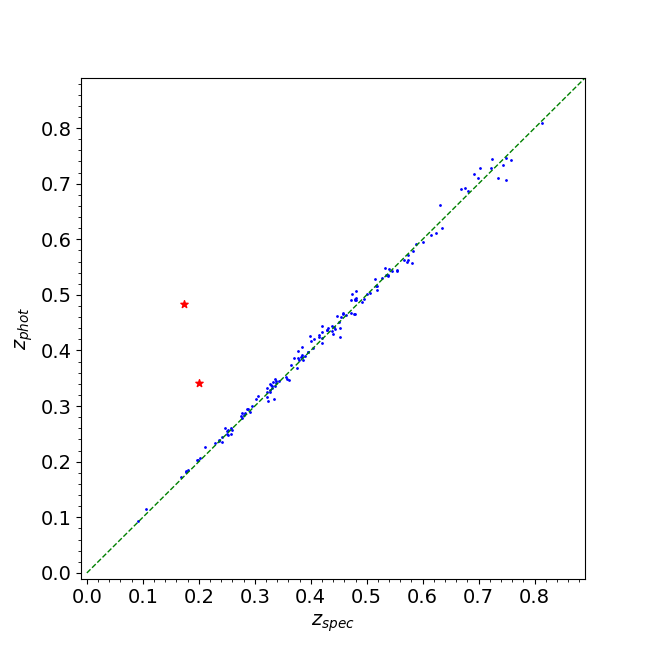}
    \caption{Relation between the photometric redshift estimates (vertical) and spectroscopic redshift measurements for the galaxy clusters from the SRG/eROSITA survey. The red stars indicate the galaxy clusters SRGe\,J144605.0$+$753727 and SRGe\,J173355.8$+$300018 with a large discrepancy between the photometric and spectroscopic redshifts.
    }
  \label{fig:zph}
\end{figure}

Redshift estimation algorithm of SRG/eROSITA galaxy clusters from the galaxies photometric redshift estimates was developed in 2023 \citep{zazn23}. Using this algorithm, we estimated the redshifts of 149 galaxy clusters at $z_{\rm spec} \lesssim 0.8$ and determined the reliability of the optical identification of clusters. Some of the galaxy clusters turned out to be outside the DESI survey region \citep{desi}, where the algorithm works. Other galaxy clusters turned out to be too distant and, hence, there are no photometric redshift estimates for the galaxies from these clusters in the catalog by \cite{zou} used by the algorithm. Therefore, the photometric algorithm estimated the redshifts of only some of the galaxy clusters. The relation between the spectroscopic redshift measurements and photometric redshift estimates is shown in fig.~\ref{fig:zph}, where the photometric estimates and spectroscopic measurements are plotted along the vertical and horizontal axes, respectively. The accuracy of the photometric estimates is $\delta z/(1+z) \approx 0.007$ for galaxy clusters at $z < 0.8$.

In the case of two galaxy clusters, SRGe\,J144605.0$+$753727 and SRGe\,J173355.8$+$300018, the algorithm provides redshift estimates that do not correspond to the spectroscopic measurements (indicated by the red stars in fig.~\ref{fig:zph}). The reliability of the optical identification of these clusters is 0.10 and 0.74, respectively, suggesting a possible inaccuracy in the identification of these galaxy clusters. For the galaxy cluster SRGe\,J173355.8$+$300018 with a reliability of 0.74 the algorithm points to the possible presence of a projection.

\subsection{Remarks on Individual Objects}

\begin{figure}
  \centering
    \includegraphics[width=0.97\columnwidth]{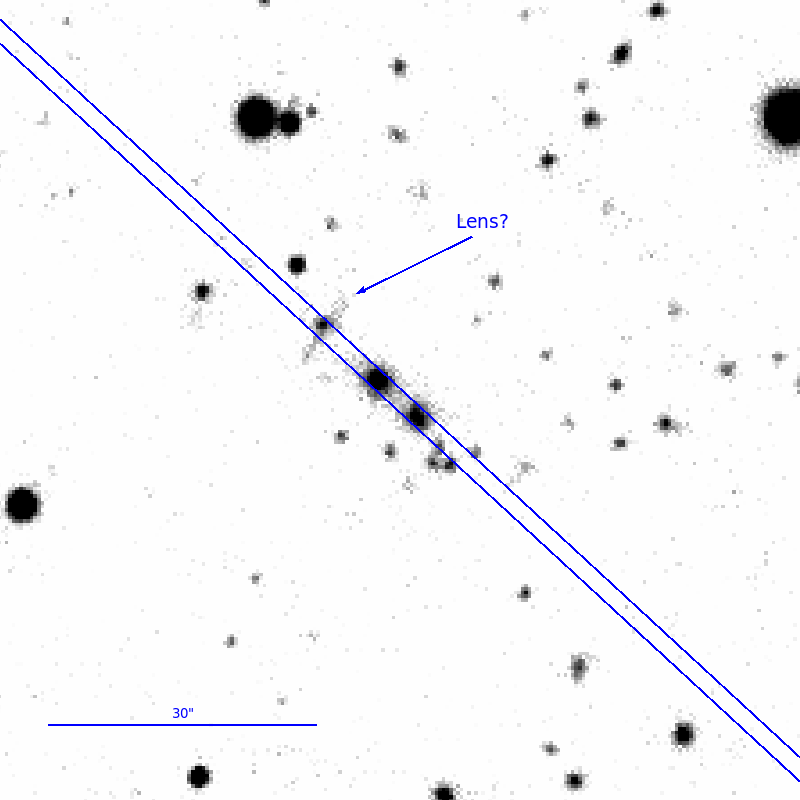}
    \caption{The image of the central region of the galaxy cluster SRGe\,J022418.2$+$301934 in the \textit{r} filter taken from DESI LIS. The blue lines indicate the contours of the TDS slit with a width of 2\arcsec. The arrow indicates the candidate for lensed galaxies.
    }
  \label{fig:cl0224:lens}
\end{figure}

{\bf SRGe\,J022418.2$+$301934.}
When inspecting the DESI LIS images, a very faint extended structure, whose shape and orientation with respect to the cluster center is typical for lensed galaxies, was detected in the central part of the cluster. An image of the candidate for lensed galaxies is presented in Fig.~\ref{fig:cl0224:lens}, the coordinates of the object are 02~24~19.6 +30~19~39. The slit configuration is presented in the figure, and the candidate for lensed galaxies is marked. It can be seen that we determined the slit position in such a way that the two brightest cluster galaxies ($r \sim 20^m$) and the candidate for lensed galaxies fell on the spectrograph slit. The observations were carried out on October 1, 2022, at the 2.5-m CMO telescope. We obtained four spectroscopic images with an exposure time of 1200~s each, the signal-to-noise ratio for the galaxies is about 2.5\,--\,2.7. The redshift of both galaxies is the same and equal to $z_{\rm spec} = 0.480$. The spectrum of the candidate for lensed galaxies is very weak and, therefore, cannot be extracted. No separate observation of this object was carried out, since this requires a large amount of observing time at big telescopes and is beyond the scope of the observing program.

{\bf SRGe\,J131253.5$+$725502.}
The effect of weak gravitational lensing of background galaxies described in \cite{dahle} is observed in this cluster. Photometric redshift estimates for the cluster are given in this paper and, therefore, we decided to carry out spectroscopic observations of this cluster at the BTA telescope.

{\bf SRGe\,J132950.1$+$564752.}
This is the most distant galaxy cluster detected in the SRG/eROSITA all-sky survey whose redshift has been measured spectroscopically. The spectrum of galaxies from this cluster was taken on May 14\,--\,15, 2021, at the BTA telescope with the SCORPIO-2 spectrograph. We obtained a total of 16 spectroscopic images with an exposure time of 1200~s each in VPHG1200@860. The coordinates of the slit center are 13~29~50.1 +56~47~52, the slit position angle is $PA = 276.7^{\circ}$. As a result of the data processing, we measured the redshift of the brightest cluster galaxy, $z_{\rm spec} = 1.298$, with coordinates 13~29~51.4 +56~47~50. The spectrum of this galaxy is presented in Fig.~\ref{fig:bta:hz} in the lower row on the right. The light from another cluster galaxy also fell on the spectrograph slit. The spectrum of this galaxy turned out to be weak and very noisy. However, the spectral features in the region 9000--9200\,\AA, which may be the Fraunhofer K and H lines, and 4000\,\AA, the jump at redshift $z \approx 1.3$, can be distinguished in it.

\begin{figure}
  \centering
    \includegraphics[width=0.97\columnwidth]{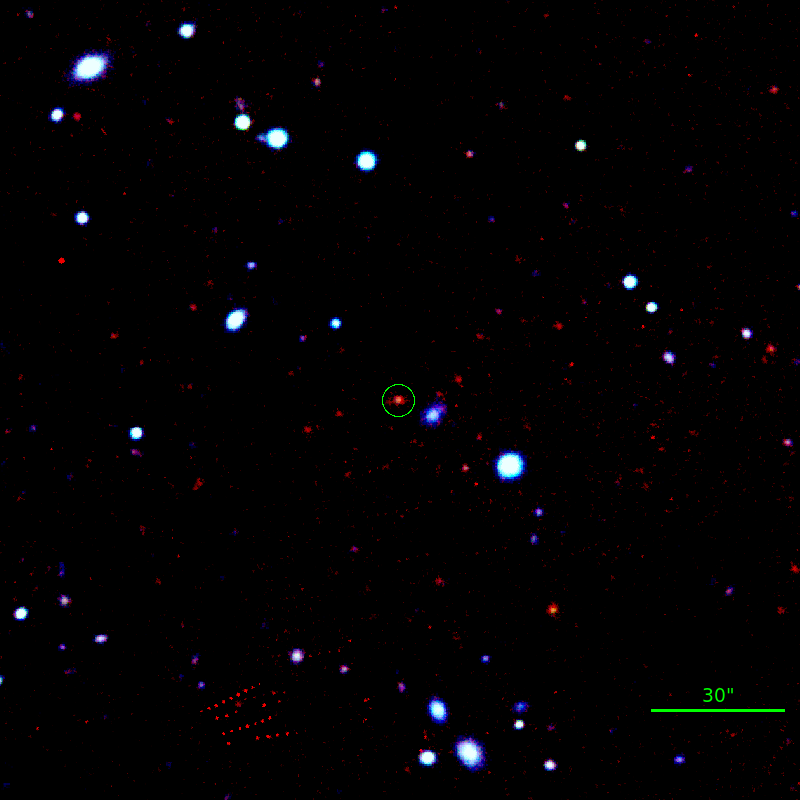}
    \caption{A pseudo-color image of the galaxy cluster SRGe\,J132950.1$+$564752 in the \textit{Jzi} (RGB) filters. The green circumference marks the brightest cluster galaxy for which the BTA spectrum was taken. The total exposure time is 1440~s in the \textit{iz} filters (BTA, SCORPIO-2) and 6945~s in the \textit{J} filter (the 2.5-m CMO telescope, AstroNIRCam).
    }
  \label{fig:cl1330izJ}
\end{figure}

This galaxy cluster is identified with the galaxy cluster MOO\,J1329$+$5647 with its photometric redshift estimate $z_{\rm phot} = 1.43 \pm 0.04$ \citep{moo}. Deep direct images were obtained for this cluster. We obtained 24 images with an exposure time of 60~s each in the \textit{iz} filters on November 9, 2021, at the BTA telescope with the SCORPIO-2 spectrograph, the seeing is 1.1\arcsec. We obtained 72 images in the \textit{J} filter with a total exposure of 6945~s at the 2.5-m CMO telescope with the AstroNIRCam instrument on December 11 and 27, 2021, and 120 images in the K filter with a total exposure time of 3500~s on December 30, 2021. A pseudo-color image of the cluster is presented in Fig.~\ref{fig:cl1330izJ}.

{\bf SRGe\,J135353.3$+$733157.}
The spectrum of the cD galaxy in this cluster contains a bright [OII]$\lambda$3727\,\AA\ line. The coordinates of the galaxy are 13~53~53.2 +73~31~56, the measured spectroscopic redshift is $z_{\rm spec} = 0.4780$. In our previous paper \citep{br22} we provided the star formation rates of the brightest galaxies in clusters estimated from the galaxy luminosity in the [OII]$\lambda$3727\,\AA\ line. For the cluster SRGe\,J135353.3$+$733157 star formation is observed in the brightest cluster galaxy in its central part. Therefore, we also estimated the star formation rate for it.

For this purpose, we estimated the interstellar extinction using the model from \cite{schlafly}, which for the cluster SRGe\,J135353.3$+$733157 with coordinates $l \approx 117^{\circ}$ and $b \approx 43^{\circ}$ turned out to be negligible, $E(B-V) = 0.007 \pm 0.025$, and it may be neglected. The ﬂux in the [OII]$\lambda$3727\AA\ line is $F_{OII\lambda3727} = (3.6 \pm 0.4) \times 10^{-15}$~erg~s$^{-1}$~cm$^{-2}$. To determine the luminosity in the line, we used the $\Lambda$CDM model of a ﬂat Universe with parameters $H_0 = 70$~km~s$^{-1}$~Mpc$^{-1}$ and $\Omega_M = 0.3$. The luminosity in the [OII]$\lambda$3727\,\AA\ line was found to be $L_{OII\lambda3727} = (3.1 \pm 0.3) \times 10^{42}$~erg~s$^{-1}$. If the possible contribution from the AGN emission is ignored, then using the relation from \cite{sfr}, the upper limit on the star formation rate can be estimated to be $SFR < 43$~M$_{\odot}$~yr$^{-1}$.

{\bf SRGe\,J143202.2$+$851737}

\begin{figure}
  \smfigure{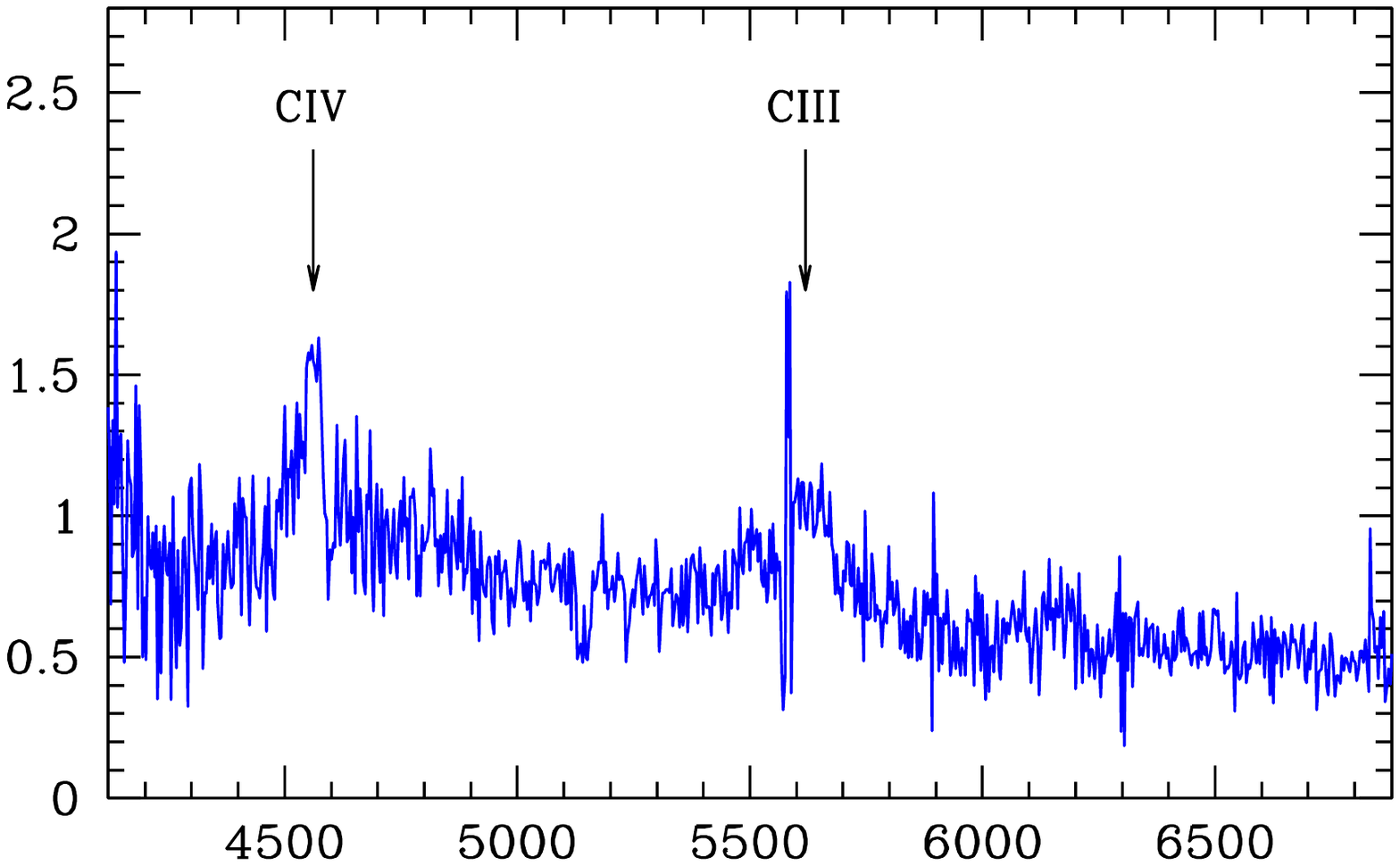}{$\lambda$, \AA}{Flux, $10^{-17}$~erg\,s$^{-1}$\,cm$^{-2}$}
  \bigskip
  
  \caption{The spectrum of the quasar at redshift $z_{\rm spec} = 1.945$ taken at the BTA telescope with the SCORPIO-2 spectrograph when observing the galaxy cluster SRGe\,J143202.2$+$851737.
  }
  \label{fig:cl1432_qso}
\end{figure}

During the BTA observations of this galaxy cluster a quasar at redshift $z_{\rm spec} = 1.945$ located at a distance of about 2\arcmin\ from the cluster center fell on the slit of the SCORPIO-2 spectrograph, the coordinates of the quasar are 14~32~03.1 +85~15~28. The redshift of the quasar was measured from the CIII$\lambda$1908\,\AA\ and CIV$\lambda$1549,\AA\ emission lines (see fig.~\ref{fig:cl1432_qso}). No X-ray source coincident in coordinates with this quasar was detected in the SRG/eROSITA survey.

\begin{figure}
  \centering
    \includegraphics[width=0.97\columnwidth]{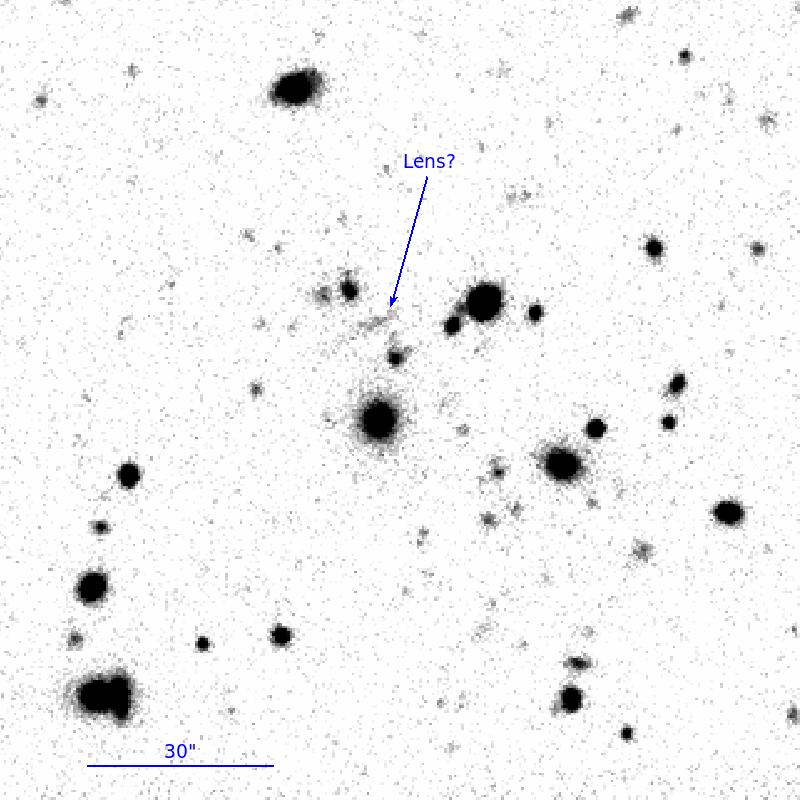}
    \caption{The image of the central region of the galaxy cluster SRGe\,J162545.1$+$672930 in the \textit{r} filter taken from DESI LIS. The arrow indicates the candidate for lensed galaxies.
    }
  \label{fig:cl1626:lens}
\end{figure}

{\bf SRGe\,J162545.1$+$672930.}
In the central region of this cluster there is a very faint extended structure in the DESI LIS images that can be a lensed galaxy, the coordinates of the object are 16~25~45.7 +67~29~46 (fig.~\ref{fig:cl1626:lens}). The brightness of the object is very low and, therefore, it was not considered as a target for our spectroscopic observations.

{\bf SRGe\,J172321.5$+$235041.}
The observations of the cluster cD galaxy with coordinates 17~23~21.6 +23~50~39 were carried out on August 3, 2022, at RTT-150. The spectrum of the galaxy is very noisy and contains the H$_{\alpha}$ and [NII]$\lambda$6584\,\AA\ emission lines and the 4000\,\AA\ jump. The H$_{\beta}$ and [OIII]$\lambda$5007\,\AA\ lines are absent at a 2$\sigma$ level. The quality of the spectrum allows the redshift of the cluster cD galaxy to be measured, but it does not allow the ﬂuxes in the H${\alpha}$ and [NII]$\lambda$6584\,\AA\ lines to be measured reliably. Therefore, to claim that there is activity in the nucleus of the cluster cD galaxy or star formation in it, it is necessary to carry out additional spectroscopic observations.

\begin{figure}
  \centering
    \includegraphics[width=0.97\columnwidth]{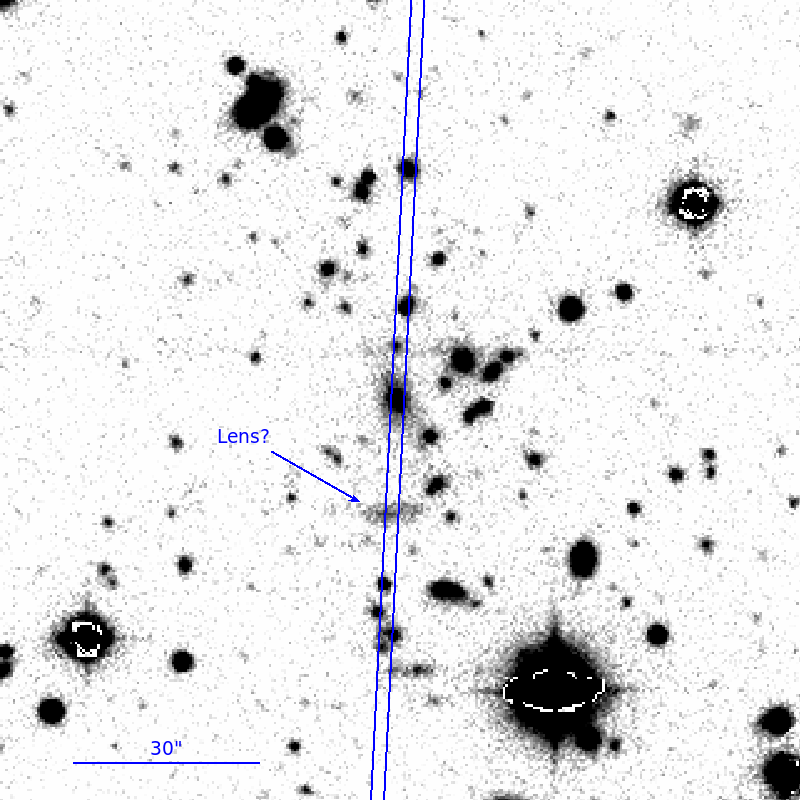}
    \caption{The image of the central region of the galaxy cluster SRGe\,J174629.3$+$472502 in the \textit{r} filter taken from DESI LIS. The blue lines indicate the contours of the 2\arcsec-wide slit. The arrow indicates the candidate for lensed galaxies.
    }
  \label{fig:cl1746:lens}
\end{figure}

{\bf SRGe\,J174629.3$+$472502.}
In the central region of this galaxy cluster there is an extended structure with coordinates 17~46~29.6 +47~24~44 that can be a lensed galaxy (Fig.~\ref{fig:cl1746:lens}). Therefore, during the observations of this galaxy cluster we used a slit configuration that allowed the spectra of a large number of cluster galaxies, including the CD galaxy, and the spectrum of the candidate for lensed galaxies to be taken. The observations of this object were carried out on June 28, 2020, at the BTA telescope with the SCORPIO-2 spectrograph. We obtained a total of three spectroscopic images with an exposure time of 900~s each at 1.7\arcsec\ seeing. We extracted the spectra of 11 objects that fell on the spectrograph slit, six of which turned out to be galaxies from the cluster SRGe\,J174629.3$+$472502 at redshift $z_{\rm spec} \approx 0.482$. The spectrum of the candidate for lensed galaxies turned out to be too weak to measure its redshift.

{\bf SRGe\,J191751.9$+$692812.}
The spectrum of galaxies from this cluster was taken on August 27, 2022, at the BTA telescope with the SCORPIO-2 spectrograph. We obtained eight spectroscopic images with an exposure time of 1200~s each in VPHG1200@860. The coordinates of the slit center are 19~17~43.9 +69~27~51, the slit position angle is $PA = 27.3^{\circ}$. As a result of the data processing we measured the redshifts of two cluster galaxies: $z_1 = 1.096 \pm 0.003$ and $z_2 = 1.101 \pm 0.005$. The sum of the spectra for these two galaxies is presented in Fig.~\ref{fig:bta:hz} in the upper row on the left.

\begin{figure}
  \centering
    \includegraphics[width=0.97\columnwidth]{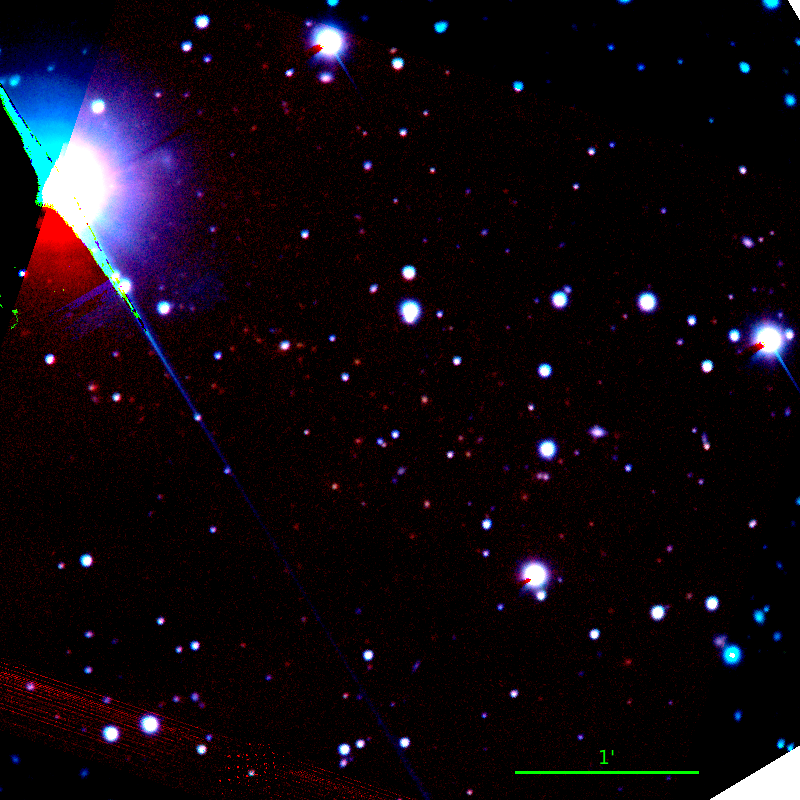}
    \caption{A pseudo-color image of the galaxy cluster SRGe\,J191751.9$+$692812 in the \textit{Jzi} (RGB) filters. The total exposure time is 540~s in the \textit{iz} filters (BTA, SCORPIO-2) and 4051~s in the \textit{J} filter (the 2.5-m CMO telescope, AstroNIRCam).
    }
  \label{fig:cl1918izJ}
\end{figure}

We also obtained deep direct images of the cluster. We obtained six images with an exposure time of 90~s each in the \textit{iz} filters at the BTA telescope with the SCORPIO-2 spectrograph and 42 images in the \textit{J} filter with a total exposure time of 4051~s at the 2.5-m CMO telescope with the AstroNIRCam instrument. A pseudo-color image of the cluster field in the \textit{Jzi} (RGB) filters is presented in Fig.~\ref{fig:cl1918izJ}.

{\bf SRGe\,J191842.1$+$744327.}
The spectrum of galaxies from this cluster was taken on October 23, 2020, at the BTA telescope with the SCORPIO-2 spectrograph. We obtained ten spectroscopic images with an exposure time of 900~s each in VPHG1200@860 at 2\arcsec\ seeing. The coordinates of the slit center are 19~18~42.1 +74~43~31, the slit position angle is $PA = 112.6^{\circ}$. As a result of the data processing, we measured the redshift of the brightest cluster galaxy, $z_{\rm spec} = 1.024$, with coordinates 19~18~45.0 +74~43~26. We failed to reliably determine the redshifts of other galaxies on the spectrograph slit. The spectrum of the galaxy is presented in fig.~\ref{fig:bta:hz} in the lower row on the right.

\begin{figure}
  \centering
    \includegraphics[width=0.97\columnwidth]{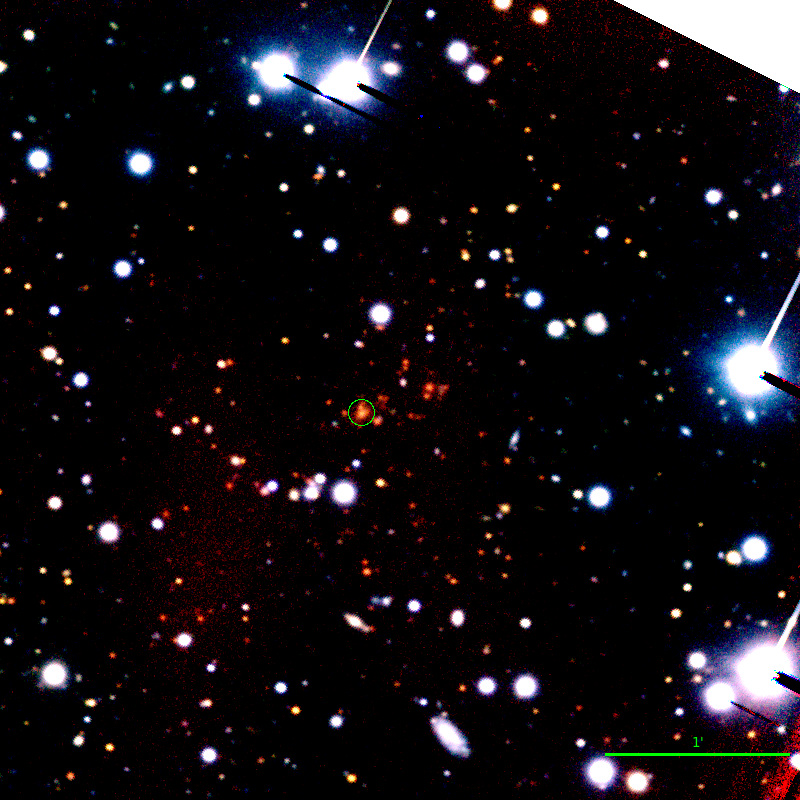}
    \caption{A pseudo-color image of the galaxy cluster SRGe\,J191842.1$+$744327 in the \textit{zir} (RGB) filters. The green circle marks the brightest cluster galaxy for which the BTA spectrum was taken. The total exposure time in the \textit{riz} filters is 1440~s (BTA, SCORPIO-2).
    }
  \label{fig:cl1919riz}
\end{figure}

On November 9, 2020, we also obtained deep direct images of the cluster. We obtained 16 images with an exposure time of 90~s each in the \textit{riz} filters at the BTA telescope with the SCORPIO-2 spectrograph. The seeing is 1.4\arcsec. A pseudo-color image of the cluster field in the \textit{zir} (RGB) filters is presented in fig.~\ref{fig:cl1919riz}.

\begin{figure}
  \centering
    \includegraphics[width=0.97\columnwidth]{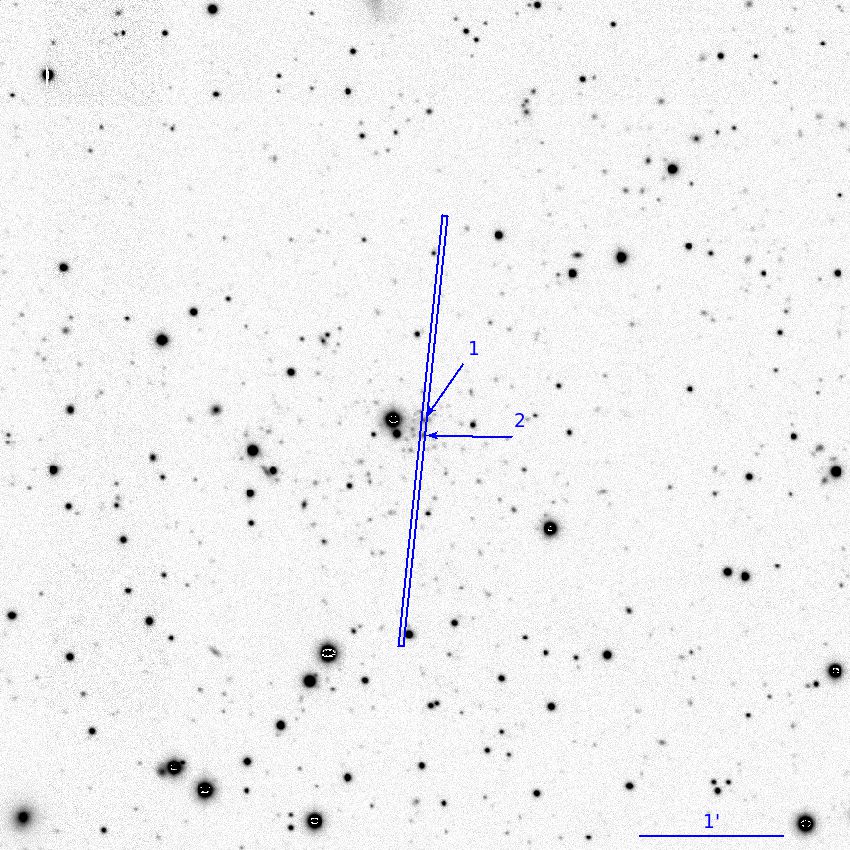}
    \caption{The pointing picture in the r filter for the galaxy cluster SRGe\,J215157.4$+$111248 for the BTA observation taken from DESI LIS. The blue rectangle indicates the orientation of the long 2\arcsec-wide slit of the SCORPIO-2 spectrograph. The slit position angle is $PA = 354.2^{\circ}$, the coordinates of the slit center are 21~51~57.9 +11~12~56. Galaxies 1 and 2 are marked by the arrows.
    }
  \label{fchart:cl2152}
\end{figure}

\begin{figure*}
  \centering
  Galaxy 1, $z = 0.924$
  \bigskip

  \smfiguresmall{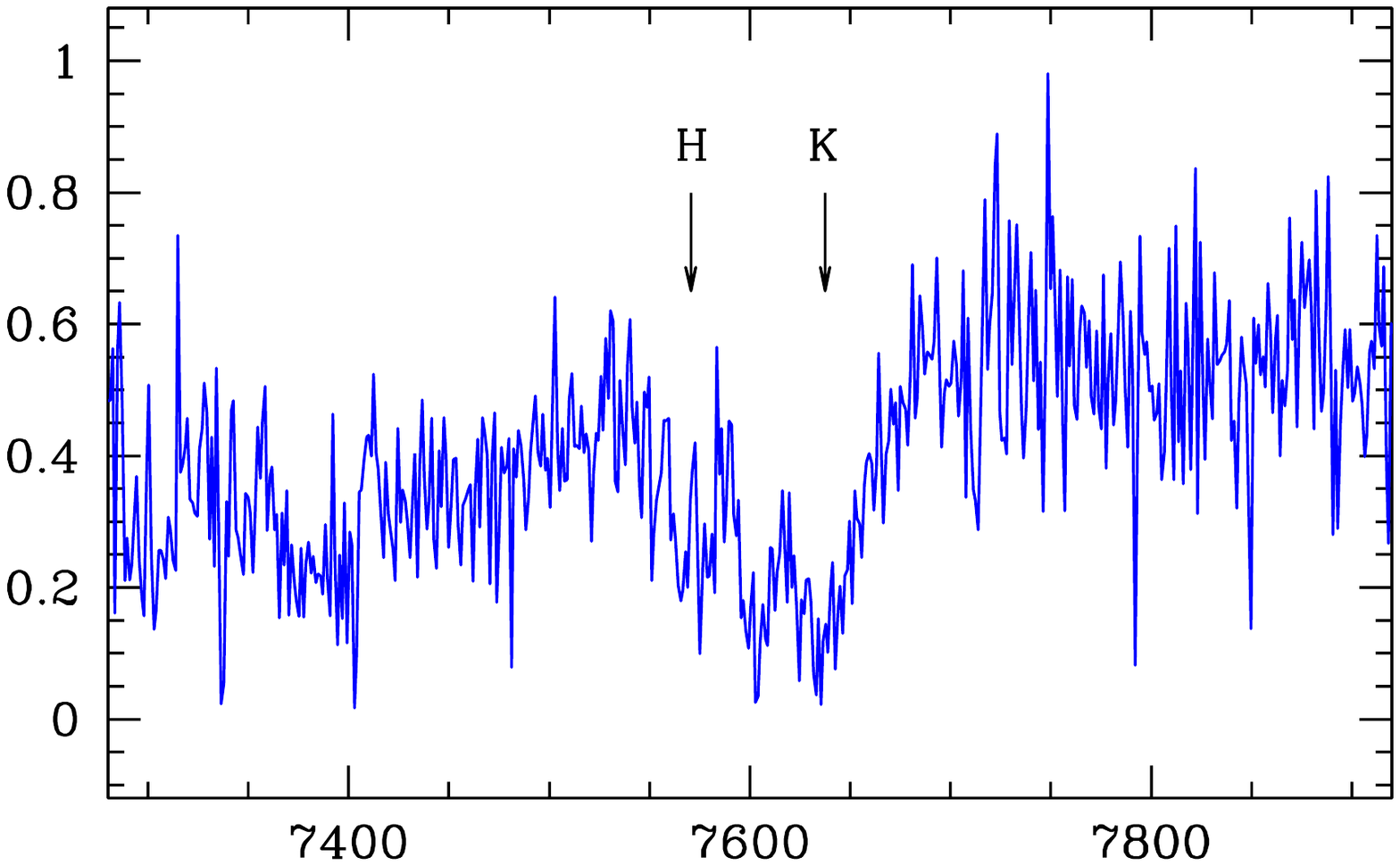}{$\lambda$, \AA}{Flux, $10^{-17}$~erg\,s$^{-1}$\,cm$^{-2}$}~\smfiguresmall{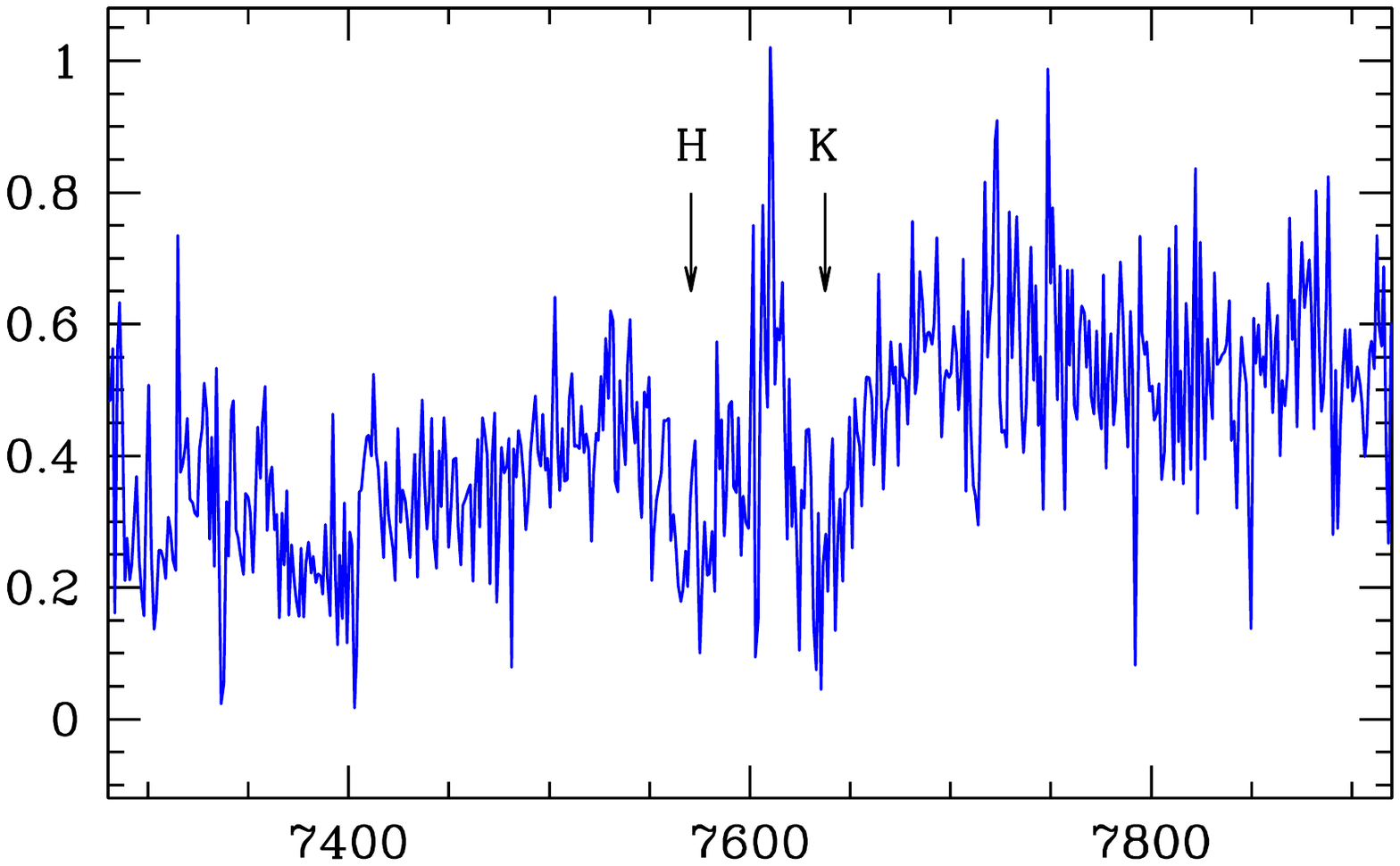}{$\lambda$, \AA}{Flux, $10^{-17}$~erg\,s$^{-1}$\,cm$^{-2}$}
  \bigskip

  \centering
  Galaxy 2, $z = 0.914$
  \bigskip
  
  \smfiguresmall{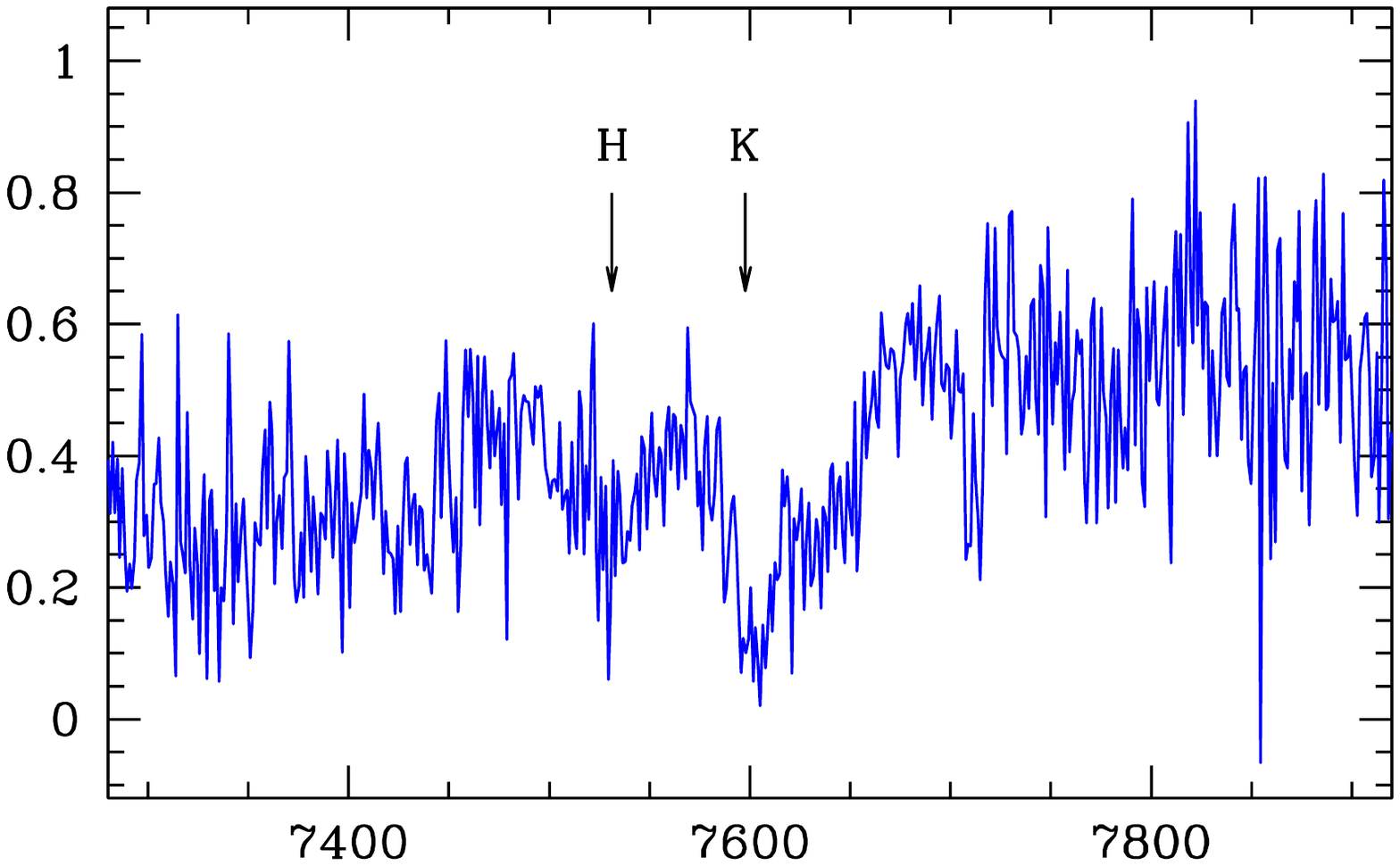}{$\lambda$, \AA}{Flux, $10^{-17}$~erg\,s$^{-1}$\,cm$^{-2}$}~\smfiguresmall{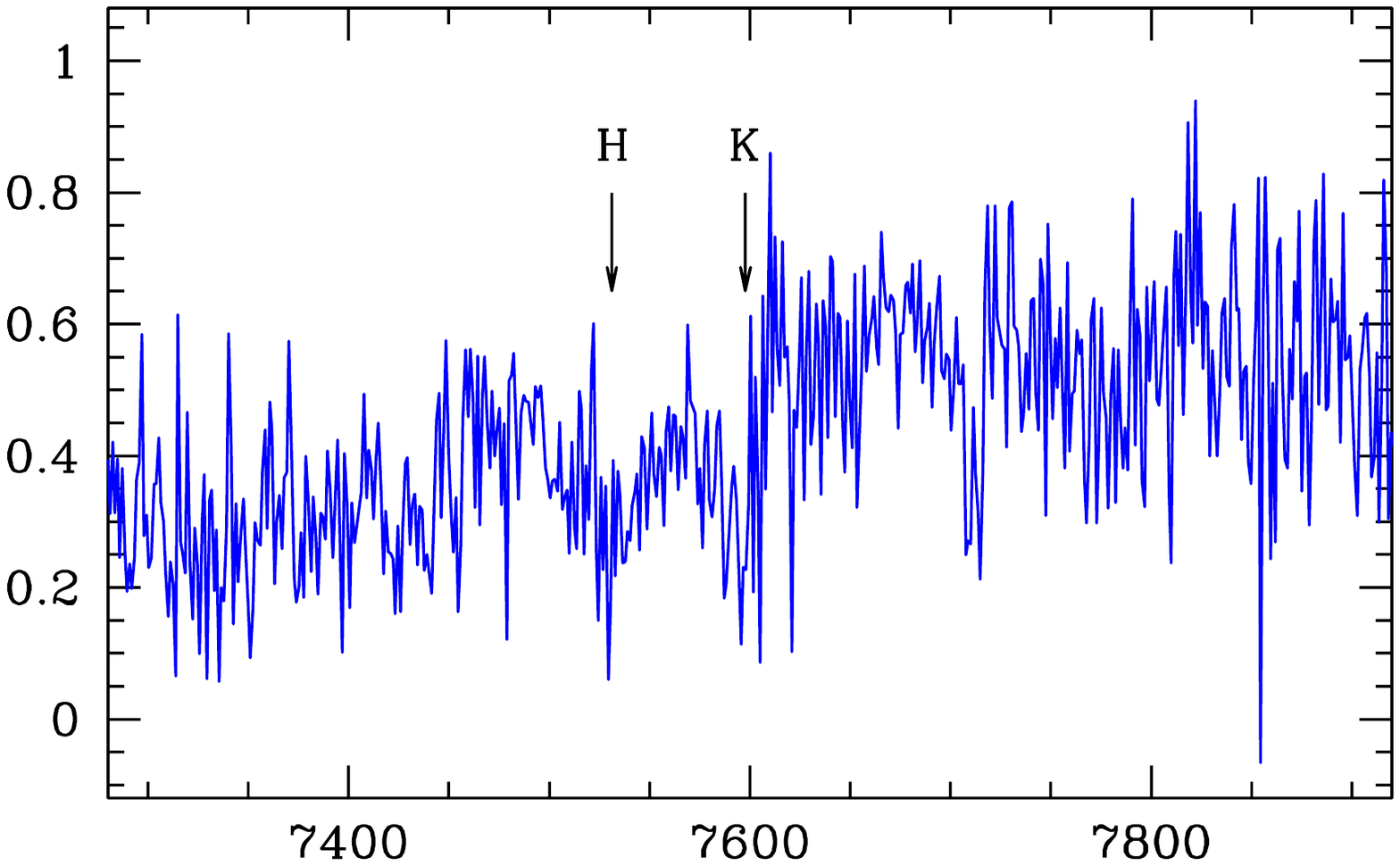}{$\lambda$, \AA}{Flux, $10^{-17}$~erg\,s$^{-1}$\,cm$^{-2}$}  
  \bigskip

  \caption{Spectra of two galaxies from the cluster SRGe\,J215157.4$+$111248. The spectra of the galaxies before and after the correction for the atmospheric $O_2$ absorption are shown in the left and right rows, respectively.
  }
  \label{fig:cl2152}
\end{figure*}

{\bf SRGe\,J215157.4$+$111248.}
For this object we took the spectra of two galaxies (Figs.~\ref{fchart:cl2152} and~\ref{fig:cl2152}), whose redshifts were estimated to be $z \sim 0.9$. Therefore, we made a correction for the atmospheric $O_2$ absorption at 7580–7700\,\AA\ and 7580--7700\,\AA\, as described above. Figure~\ref{fig:cl2152} presents the spectra of the galaxies before and after the correction for the atmospheric absorption. After the correction for the absorption we detected the Fraunhofer H and K absorption lines in the spectra of the galaxies. A comparison of their spectra with the spectrum of a template with an age of 2.5~Gyr and $Z = 0.008$ allows the redshifts of the galaxies to be determined as $z_{\rm spec} = 0.914$ and $z_{\rm spec} = 0.924$. Therefore, we determined the redshift of the cluster SRGe\,J215157.4$+$111248 as $z = 0.919$.

\section{CONCLUSIONS}

From June 2020 to September 2023 we obtain the spectra and measured the redshifts for 216 galaxy clusters as a result of the implementation of our program of optical identifications and spectroscopic observations of the galaxy clusters detected during the SRG/eROSITA all-sky survey. Among them 139 galaxy clusters have been discovered for the first time. In total, we obtain the spectra of 106, 45, 51, and 26 galaxy clusters at the BTA, 2.5-m CMO, AZT-33IK, and RTT-150 telescopes, respectively.

In total, we measured the redshifts of 22 distant galaxy clusters at $z > 0.7$. The most distant galaxy cluster detected during the SRG/eROSITA all-sky survey whose spectroscopic redshift was measured is at redshift $z_{\rm spec} = 1.298$. For some of the distant galaxy clusters we obtained deep direct images at the BTA, AZT-33IK, and RTT-150 telescopes in the \textit{riz} SDSS filters and in the \textit{J} filter at the 2.5-m CMO telescope. In future, we will continue this work to obtain calibrations the dependence of the red-sequence color for distant ($z > 0.7$) galaxy clusters on their redshifts. In the brightest galaxies of some clusters we detected emission lines that are traces of star formation in the galactic nucleus. When inspecting the direct images from DESI LIS for the fields of galaxy clusters, we detected several candidates for lensed galaxies.

The presented data are the result of our three year work on optical identifications and spectroscopic redshift measurements of some of the most massive galaxy clusters detected in the SRG/eROSITA all-sky survey. It is expected that, in future, the work on optical identifications and spectroscopic measurements of galaxy clusters from the SRG/eROSITA all-sky survey will be continued. We are also going to continue the photometric observations of distant galaxy clusters located at redshifts $z\sim1$ or higher for their optical identification and the subsequent calibration of the photometric redshift estimates for distant galaxy clusters from the color of their red sequences.

\acknowledgements

This work was supported by RSF grant no. 21-12-00210. The observations at the SAO RAS telescopes are supported by the Ministry of Science and Higher Education of the Russian Federation. The instrumentation is updated within the National Project “Science and Universities”. The work of SD, SK, EM, AM, DO, RU, and ESh to obtain the observational data was performed within the State assignment of SAO RAS approved by the Ministry of Science and Higher Education of the Russian Federation. The work of ME was supported by the Ministry of Education and Science of Russia. The AZT-33IK results were obtained using the equipment of the Angara sharing center (http://ckp-rf.ru/ckp/3056/). This work was supported in part by the Program for the Advancement of the Moscow State University (the scientific and educational school “Fundamental and Applied Space Research”). We are grateful to T\"{U}B\.{I}TAK, the Space Research Institute, the Kazan Federal University, and the Academy of Sciences of Tatarstan for their partial support in using RTT-150 (the Russian–Turkish 1.5-m telescope in Antalya). The work of IKh, EI, and NS was supported by subsidy no. FZSM-2023-0015 of the Ministry of Education and Science of the Russian Federation allocated to the Kazan Federal University for the State assignment in the sphere of scientific activities. In this study we used observational data from the eROSITA telescope onboard the SRG observatory. The SRG observatory was built by Roskosmos in the interests of the Russian Academy of Sciences represented by the Space Research Institute within the framework of the Russian Federal Space Program, with the participation of the Deutsches Zentrum f\"{u}r Luft- und Raumfahrt (DLR). The SRG spacecraft was designed, built, launched, and is operated by the Lavochkin Association and its subcontractors. The science data are downlinked via the Deep Space Network Antennae in Bear Lakes, Ussurijsk, and Baykonur, funded by Roskosmos. The SRG/eROSITA X-ray telescope was built by a consortium of German Institutes led by MPE, and supported by DLR. The eROSITA data used in this work were processed using the eSASS software developed by the German eROSITA consortium and the proprietary data reduction and analysis software developed by the Russian eROSITA Consortium. In this study we used the NASA/IPAC Extragalactic Database (NED) operated by the Jet Propulsion Laboratory of the California Institute of Technology under contract with the National Aeronautics and Space Administration. In this paper we used the data from the photometric DESI survey obtained at the Blanco Telescope of the Cerro Tololo Inter-American Observatory, the Bok Telescope of the Steward Observatory of the University of Arizona, and the Mayall Telescope of the Kitt Peak National Observatory.

\bibliographystyle{pazh}

\bibliography{refs_rus}

\end{document}